\documentclass[preprint,11pt, 3p, authoryear]{elsarticle}
\usepackage{mathtools}
\usepackage{esint}
\usepackage{graphicx}
\usepackage{graphics}
\usepackage{multirow}
\usepackage{bm}
\usepackage{listings}
\usepackage{color}
\usepackage[colorlinks]{hyperref}
\usepackage{cancel}
\usepackage{braket}
\usepackage{relsize}
\usepackage{bm}
\usepackage{amsmath}
\usepackage{amsbsy}
\usepackage{amssymb}
\usepackage{amsfonts}
\usepackage{mathrsfs}
\usepackage{sidecap}
\usepackage{subcaption}
\DeclareMathOperator{\diag}{\mathit{diag}}

\journal{arXiv}

\def \bx{\mathbf{x}}
\def \bR{\mathbf{R}}
\def \E{E}
\def \I{a}
\def \N{N_A}
\def \hN{\hat{N}_A}
\def \t{t}
\def \Nt{N_T}
\def \wt{w_t}
\def \k{k}
\def \Z{Z}
\def \bX{\mathbf{X}}
\def \bXt{\mathbf{\tilde{X}}}
\def \l{l}
\def \L{N_L}
\def \m{m}
\def \e{e}
\def \Ne{N_E}
\def \n{n}
\def \Nr{N_R}
\def \J{b}
\def \hnl{h_{nl}}
\def \Ylm{Y_{lm}}
\def \bC{\mathbf{C}}
\def \cnlm{\varrho_{nlm}}
\def \bRJI{\mathbf{R}_b - \mathbf{R}_a}

\def \K{c}
\def \f{f}
\def \bf{\mathbf{f}}
\def \R{R}
\def \bXIt{\mathbf{W}_{a,t}}

\def \M{M}
\def \Q{d}
\def \sfgm{\Gamma_{\zeta, \mu}}
\def \sfgmb{\Gamma_{\bar{\zeta}, \bar{\mu}}}
\def \sfgmt{\Gamma_{\tilde{\zeta}, \tilde{\mu}}}
\def \sfgmh{\Gamma_{\hat{\zeta}, \hat{\mu}}}
\def \rotM{\mathbf{Q}^{\zeta}_{\Theta} \mathbf{Q}^{\mu}_\varphi}
\def \rotMinv{\mathbf{Q}^{-\zeta}_{\Theta} \mathbf{Q}^{-\mu}_\varphi}
\def \rotMt{\mathbf{Q}^{\tilde{\zeta}}_{\Theta} \mathbf{Q}^{\tilde{\mu}}_\varphi}

\def \rotMh{\mathbf{Q}^{\hat{\zeta}}_{\Theta} \mathbf{Q}^{\hat{\mu}}_\varphi}
\def \rotMhinv{\mathbf{Q}^{-\hat{\zeta}}_{\Theta} \mathbf{Q}^{-\hat{\mu}}_\varphi}
\def \rotMMhinv{\mathbf{Q}^{-(\zeta+\hat{\zeta})}_{\Theta} \mathbf{Q}^{-(\mu+\hat{\mu})}_\varphi}
\def \rotMMtinv{\mathbf{Q}^{-(\tilde{\zeta}+\hat{\zeta})}_{\Theta} \mathbf{Q}^{-(\tilde{\mu}+\hat{\mu})}_\varphi}
\def \tmu{\mu \tau \mathbf{\hat{e}_3}}  %{\mathbf{t}_\mu}
\def \bD{\mathbf{D}}
\def \bu{\mathbf{u}}
\def \bq{\mathbf{q}}
\def \sfG{\mathcal{G}}
\def \sfGp{\mathcal{\bar{G}}}
\def \bI{\mathbf{I}}

\def \C{C_{enlm}}
\def \bw{\mathbf{w}}
\def \wE{w_E}
\def \wf{w_f}
\def \yE{\mathbf{y}_E}
\def \yf{\mathbf{y}_f}
\def \yEt{\mathbf{\tilde{y}}_E}
\def \yft{\mathbf{\tilde{y}}_f}
\def \KE{\mathbf{K}_E}
\def \Kf{\mathbf{K}_f}
\def \KEt{\mathbf{\tilde{K}}_E}
\def \Kft{\mathbf{\tilde{K}}_f}
\def \sigE{\sigma_E}
\def \sigf{\sigma_f}
\def \Nst{N_{st}}
\def \sigy2s{\mathbf{\sigma}^2_{\mathbf{y}_s}}

\makeatletter
\newcommand*{\rom}[1]{\expandafter\@slowromancap\romannumeral #1@}
\makeatother

\begin{document}
%%%%%%%%%%%%%%%%%%%%%%%%%%%%%%%%%%%%%%%%%%%%%%%%%%%%%%%%%%%%%%%%%%%%%%%%%
\begin{frontmatter}
\title{Cyclic and helical symmetry-informed machine learned force fields: Application to lattice vibrations in carbon nanotubes}

\author[llnl]{Abhiraj Sharma\textsuperscript{1,}}
\author[gatech_coe]{Shashikant Kumar\textsuperscript{1,}}       
\author[gatech_coe,gatech_coc]{Phanish Suryanarayana\corref{cor}}

%-------------------------
\address[llnl]{Physics Division, Lawrence Livermore National Laboratory, Livermore, CA 94550, USA}
\address[gatech_coe]{College of Engineering, Georgia Institute of Technology, Atlanta, GA 30332, USA}
\address[gatech_coc]{College of Computing, Georgia Institute of Technology, Atlanta, GA 30332, USA}

\cortext[cor]{Corresponding Author (\it phanish.suryanarayana@ce.gatech.edu) }
\fntext[contrib]{Equal contribution.}
%%%%%%%%%%%%%%%%%%%%%%%%%%%%%%%%%%%%%%%%%%%%%%%%%%%%%%%%%%%%%%%%%%%%%%%%%
\begin{abstract}
We present a formalism for developing cyclic and helical symmetry-informed machine learned force fields (MLFFs). In particular, employing the smooth overlap of atomic positions descriptors with the polynomial kernel method, we derive cyclic and helical symmetry-adapted expressions for the energy, atomic forces, and phonons (describe lattice vibration frequencies and modes). We use this formulation to construct a symmetry-informed MLFF for carbon nanotubes (CNTs), where the model is trained through Bayesian linear regression, with the data generated from ab initio density functional theory (DFT) calculations performed during on-the-fly symmetry-informed MLFF molecular dynamics simulations of representative CNTs. We demonstrate the accuracy of the MLFF model by comparisons with DFT calculations for the energies and forces, and density functional perturbation theory calculations for the phonons, while considering CNTs not used in the training. In particular, we obtain a root mean square error of $1.4 \times 10^{-4}$ Ha/atom, $4.7 \times 10^{-4}$ Ha/Bohr, and 4.8 cm$^{-1}$ in the energy, forces, and phonon frequencies, respectively, which are well within the accuracy targeted in ab initio calculations. We apply this framework to study phonons in CNTs of various diameters and chiralities, where we identify the torsional rigid body mode that is unique to cylindrical structures and establish laws for variation of the phonon frequencies associated with the ring modes and radial breathing modes. Overall, the proposed formalism provides an avenue for studying nanostructures with cyclic and helical symmetry at ab initio accuracy, while providing orders-of-magnitude speedup relative to such methods. 
\end{abstract}

%\maketitle
%\allowdisplaybreaks

\end{frontmatter}
%%%%%%%%%%%%%%%%%%%%%%%%%%%%%%%%%%%%%%%%%%%%%%%%%%%%%%%%%%%%%%%%%%%%%%%%%%
\section{Introduction}

Nanostructures represent an arrangement of atoms confined to the nanometer scale along one or more of the spatial dimensions. These atomic arrangements are not just limited to engineered structures such as nanosheets, nanotubes, nanorods, nanowires, nanoribbons, nanodots, and nanoclusters, but are also naturally occurring, such as DNA, viruses, and proteins. Over the past three decades, nanostructures have been the focus of intensive research due to their intriguing mechanical, vibrational, thermal, electronic, and optical properties \citep{xia2003one}. Among the different symmorphic and non-symmorphic nanostructures, those with cyclic and helical symmetries are perhaps be the most prevalent \citep{james2006objective, allen2007nanocrystalline1}. In addition, deformation mechanisms such as bending in the case of 1D and 2D nanostructures \citep{james2006objective, Koskinen2010RPBC}, and twisting in the case of 1D nanostructures \citep{james2006objective, Koskinen2010RPBC}, frequently give rise to cyclic and helical symmetries, even if not  present in the undeformed nanostructure. 

The development of symmetry-adapted methods are of notable interest since they not only enable significant reduction in the computational cost of the associated calculation, but also offer substantial simplification in the analysis of the results, particularly in the case of cyclic and helical symmetries.  This has motivated the development of a range of  computational methods  that exploit the cyclic and helical symmetry in nanostructures, ranging from atomistic force fields \citep{james2006objective, DumitricaobjectiveMD, DayalnonequilibriumMD, aghaei2011symmetry,  aghaei2013symmetry}  to tight-binding \citep{mintmire1993symmetries, allen2007nanocrystalline3, allen2007nanocrystalline2, popov2004carbon, popov2006radius, gunlycke2008lattice, Zhang2009CNT, Zhang2009dislocation, Koskinen2010RPBC, zhang2017inhomogeneous} to first-principles Kohn-Sham density functional theory (DFT) \citep{saito1992electronic, OnoHir2005, OnoHir2005gold, Banerjee2016cyclic, ghosh2019symmetry, banerjee2021ab, sharma2021real}. On the one hand, atomistic calculations are highly  efficient, however their accuracy and predictibility is still limited, owing to the reliance on empirical and/or semi-empirical interatomic potentials. On the other hand, though highly accurate,  the computational cost associated with DFT calculations remains particularly high even with symmetry-adaption, and becomes prohibitive for calculations based on density functional perturbation theory (DFPT) \citep{baroni1987green, gonze1989density, sharma2023calculation}, required for studying the lattice vibrations, also referred to as phonons in the literature. 

Machine-learned models that have been trained to high-fidelity ab initio data  have recently emerged as viable alternatives to the aforementioned methods. In recent work, a cyclic and helical symmetry-adapted machine-learned surrogate model has been developed for predicting the electron and pseudocharge densities in carbon nanotubes (CNTs) \citep{pathrudkar2022machine}. However, obtaining the ground state energy and atomic forces within this framework requires a non self-consistent Kohn-Sham calculation, which is still computationally expensive and cubic scaling with system size. Moreover, the cost associated with the calculation of the phonons remains prohibitively high. Machine-learned force fields (MLFFs) \citep{manzhos2006random, behler2007generalized, bartok2010gaussian, rupp2012fast, behler2015constructing, thompson2015spectral, schutt2018schnet, deringer2019machine, drautz2019atomic, von2020exploring, von2020retrospective, keith2021combining, batzner20223, batatia2022mace, ko2023recent, merchant2023scaling} provide an attractive option for the calculation of energy, forces, and phonons for atomic systems.  In particular, MLFFs can have high accuracy-to-cost ratios, which allows they to achieve near quantum mechanical accuracy in modeling material properties at length and time scales that approach experiments \citep{unke2021machine, botu2017machine, chmiela2018towards, wu2023applications}. MLFFs are typically designed to preserve the relationship between the structures and their properties. This can be effectively achieved by learning the symmetries of the underlying physics \citep{behler2007generalized, bartok2010gaussian, rupp2012fast, musil2021physics}. In general, this learning improves with the amount of training data available, however, learning the global symmetries like those found in nanostructures may require generation of a large amount of high-fidelity training data, which might be impractical. Moreover, as discussed previously, symmetry-adaption of MLFF provides a convenient avenue to reduce the computational cost of predictions and enhance the interpretability of results. In the context of nanostructures with cyclic and helical symmetries, this provides the motivation to develop cyclic and helical symmetry-informed MLFFs, which has not been targeted heretofore. 

In this work, we present a formalism for developing cyclic and helical symmetry-informed MLFFs. In particular, employing the smooth overlap of atomic positions (SOAP) descriptors along with the polynomial kernel method, we derive cyclic and helical symmetry-adapted expressions for the energy, atomic forces, and phonons. We use this formulation to construct a symmetry-informed MLFF for single-walled CNTs, where the model is trained through Bayesian linear regression, the corresponding data generated from DFT calculations performed during on-the-fly symmetry-informed MLFF molecular dynamics (MD) simulations of representative CNTs. We demonstrate the accuracy of the MLFF model by comparisons with DFT calculations for the energies and forces, and DFPT calculations for the phonons, while considering CNTs not used in the training. We apply this framework to study phonons in CNTs of various diameters and chiralities.

The remainder of this manuscript is organized as follows. In Section~\ref{Sec:nanostructure}, we provide a mathematical atomic representation for nanostructures with cyclic and helical symmetry. In Section~\ref{Sec:SIMLFF}, we discuss the cyclic and helical symmetry-informed MLFF scheme.  In Section~\ref{Sec:Results}, we verify the accuracy and efficiency of the scheme and use it to study phonons in CNTs. Finally, we provide concluding remarks in Section~\ref{Sec:ConcludingRemarks}.
 
%%%%%%%%%%%%%%%%%%%%%%%%%%%%%%%%%%%%%%%%%%%%%%%%%%%%%%%%%%%%%%%%%%%%
\section{Cyclic and helical symmetry-based atomic representation} \label{Sec:nanostructure}
\begin{figure}[htbp!]
\centering
\includegraphics[keepaspectratio=true,width=0.23\textwidth]{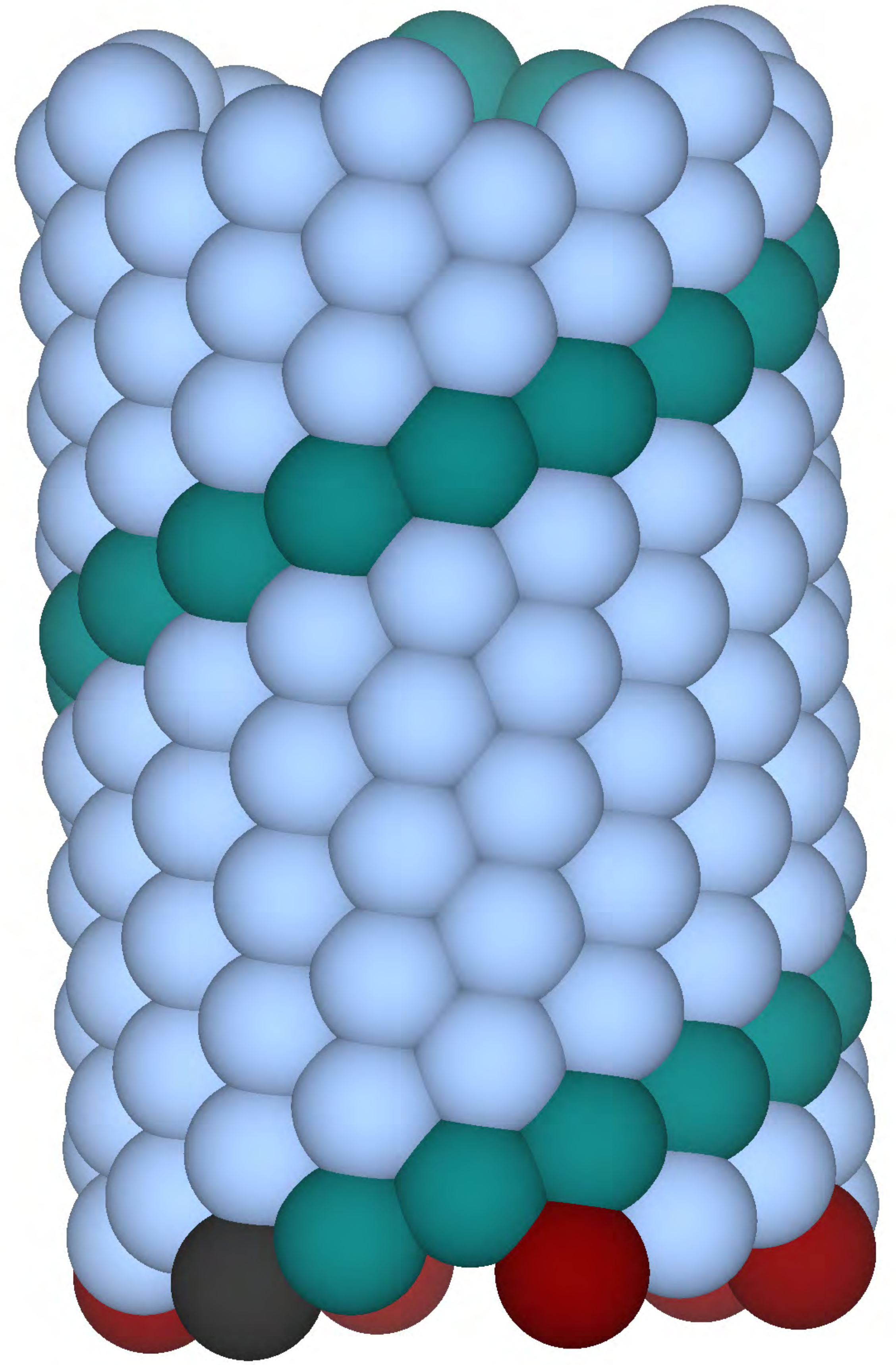}
\caption{\label{Fig:nanostructure} Illustration of a nanostructure with cyclic and helical symmetry. The fundamental atom is colored black and its cyclic and helical images are colored red and green, respectively. All other atoms within the nanostructure are the cyclic and/or helical images of these colored atoms.}
\end{figure}

Consider an infinitely long nanostructure with cyclic/rotational symmetry in the plane perpendicular to its length and helical/screw axis symmetry along its length, as illustrated in Fig.~\ref{Fig:nanostructure}. Since the action of a cyclic and helical symmetry group preserves the chemical identity of an atom \textemdash characterized by properties such as atomic number, mass number, and electronegativity\textemdash  while transforming its atomic coordinates, the nanostructure can be represented as an orbit\footnote{The orbit of an atom under the action of a symmetry group is the set of all points in Euclidean space to which the atom is moved by the elements of the group. These points are referred to as the images of the atom.} of its fundamental atoms (also referred to as motif) under the action (denoted by $\circ$) of the associated cyclic and helical symmetry group:
\begin{align}
\bR = \bigcup_{\I=1}^{\hN} \sfG \circ \bR_\I\ = \bigcup_{\I=1}^{\hN} \{\sfgm \circ \bR_\I = \rotM \bR_\I +\tmu : \zeta = 0,1,2, \ldots, \mathfrak{N}-1, \mu \in \mathbb{Z} \}
\end{align}
where $\bR = \{\bR_1,\bR_2, \ldots\}$ is the set of atomic coordinates in Euclidean space, $\hN$ is the number of fundamental atoms in the nanostructure, $\sfG$ is the associated cyclic and helical symmetry group expressed as the direct product of the cyclic symmetry group $\mathscr{C}$ and the helical symmetry group $\mathscr{S}$.  These groups are constructed by the isometries as:
\begin{align}
\mathscr{C} = \{(\mathbf{Q}^{\zeta}_{\Theta}|0 \mathbf{\hat{e}_3}): \zeta = 0,1,2, \ldots, \mathfrak{N}-1\}, \, \mathscr{S} = \{(\mathbf{Q}^{\mu}_\phi|\tmu): \mu \in \mathbb{Z}\} \,,
\end{align}
where $\Theta = \frac{2\pi}{\mathfrak{N}}$ is the angle in the plane subtended by the fundamental atom and its nearest cyclic image, $\phi$ is the angle in the plane subtended by the fundamental atom and its nearest helical image that is separated by $\tau$ along the length of the nanostructure, and
\begin{align}
\mathbf{Q}_\Theta =
\begin{bmatrix}
\cos \Theta   & -\sin \Theta & 0 \vspace{0.05in}\\
\sin \Theta & \cos \Theta & 0 \vspace{0.05in} \\
0 & 0 & 1
\end{bmatrix}  \in SO(3) \,, \,
\mathbf{Q}_\phi =
\begin{bmatrix}
\cos \phi   & -\sin \phi & 0 \vspace{0.05in}\\
\sin \phi & \cos \phi & 0 \vspace{0.05in} \\
0 & 0 & 1
\end{bmatrix} \in SO(3) \,, \,
\mathbf{\hat{e}}_3 = 
\begin{bmatrix}
0 \vspace{0.05in}\\
0 \vspace{0.05in} \\
1 
\end{bmatrix} \,,
\end{align}
with $SO(3)$ being the special orthogonal group consisting of all rotations about the origin in three-dimensional Euclidean space. This provides a unique representation to the nanostructure in terms of its fundamental atoms and the associated cyclic and helical symmetry group. 

Since the entire nanostructure can be generated by orthogonal transformations and translations of the fundamental atoms, it can be considered to be an objective structure \citep{james2006objective}, where image atoms see the same environment upto a rotation and translation. Also note that we henceforth work within a large substructure of the aforedescribed nanostructure consisting of $\N$ atoms and generated by $\sfGp =  \bigg\{(\mathbf{Q}^{\zeta}_{\Theta}|0 \mathbf{\hat{e}_3})(\mathbf{Q}^{\mu}_{\phi}|\tmu): \zeta = 0,1,2, \ldots, \mathfrak{N}-1, \mu = -\frac{1}{2}\left(\frac{\N}{\hN \mathfrak{N}} - 1\right), -\frac{1}{2}\left(\frac{\N}{\hN \mathfrak{N}} - 1\right) + 1, \ldots ,-1, 0, 1, \ldots, \frac{1}{2}\left(\frac{\N}{\hN \mathfrak{N}} - 1\right)\bigg\} \  \subseteq \sfG$, to avoid dealing with mathematical infinities in extensive quantities. This comes without any loss of generality and since the environment around the atoms in the substructure remains invariant it allows for symmetry-adaption of the quantities of interest.  We refer to this substructure as the nanostructure in the remainder of this paper, unless otherwise stated.

%%%%%%%%%%%%%%%%%%%%%%%%%%%%%%%%%%%%%%%%%%%%%%%%%%%%%%%%%%%%%%%%%%%%%%%%%%%%%%%%%%%%%%%%%%%%%%%%%%%%%%%%%%%%%%%%%%
\section{Cyclic and helical symmetry-informed MLFF scheme} \label{Sec:SIMLFF}
In this section, we present the cyclic and helical symmetry-informed formulation for MLFFs within the SOAP framework. In particular, we first describe the SOAP model for the energy, forces, and phonons in Section~\ref{Subsec:SOAP}. Next,  exploiting the cyclic and helical symmetry in the nanostructure, we derive the symmetry-adapted version of the model in Section~\ref{Subsec:SOAP:Symmetry}. Finally, we discuss the calculation  of the  model weights in Section~\ref{Subsec:Training}. 

%%%%%%%%%%%%%%%%%%%%%%%%%%%%%%%%%%%%%%%%%%%%%%%%%%%%%%%%%%%%%%%%%%%%%%%%%
\subsection{SOAP model} \label{Subsec:SOAP}
In the SOAP model, the total energy of the nanostructure is decomposed as follows:
\begin{align}\label{Eqn:Energy_mlff_total}
\E = \sum_{\I=1}^{\N} \E_\I = \sum_{\I=1}^{\N} \sum_{\t=1}^{\Nt} \wt \k(\bX_{\I},\bXt_{\t}) \,,
\end{align}
where $\E_\I$ is the atomic energy, $\wt$ is the training weight, $\bX_\I$ is the atomic descriptor corresponding to atom $\I$ in the nanostructure, $\bXt_\t$ is an element of the set which consists of $\Nt$ training descriptors\footnote{The set of training descriptors consists of atomic descriptors corresponding to each chemical species in the nanostructure, simulating the spatial and chemical environments similar to that found in the nanostructure and generated using the same mapping scheme as that employed for the nanostructure.}, and $\k$ is the similiarity kernel, for which we employ the polynomial variant:
\begin{align} \label{Eqn:kernel}
\k(\bX_{\I},\bXt_{\t}) = \delta_{\Z_\I \Z_\t} \left(\frac{\bX_\I \cdot \bXt_\t}{|\bX_\I| |\bXt_\t|}  \right)^\xi \,.
\end{align}
Above, $\delta_{\I\J}$ is the Kronecker delta function, $\Z_\I$ is the atomic number of atom $\I$,  and $\xi$ is the hyperparameter to control the sharpness of the kernel. These atomic descriptors, which are generated by a symmetry and permutation invariant smooth mapping from structural space to feature space \citep{musil2021physics}, take the following atom-centered localized form within the 3-body SOAP framework:
\begin{align}\label{Eqn:descriptor}
\bX_{\I} &= \diag \left( \mathlarger{\bigoplus}_{\l=0}^\L \sqrt{\frac{8 \pi^2}{2\l+1}} \sum_{\m=-\l}^\l \left(\bC_{\I,\l, \m} \otimes \bC^{*}_{\I,\l,\m}\right) \right) \,,
\end{align} 
where
\begin{align}\label{Eqn:C}
\bC_{\I,\l,\m} &= \mathlarger{\mathlarger{\bigoplus}}_{\e=1}^{\Ne} \mathlarger{\mathlarger{\bigoplus}}_{\n=1}^{\Nr} \left(\sideset{}{'} \sum_{\J} \delta_{\Z_\e \Z_\J} \cnlm(\bR_\J -\bR_{\I})\right) \,.
\end{align}
Above, $\diag (\cdot)$ represents a column vector constructed from the diagonal elements of the matrix, $\oplus$ denotes direct sum of the matrices, $\otimes$ denotes Kronecker product of the matrices, $\sideset{}{'} \sum$ indicates exclusion of $\bR_\J = \bR_\I$ term in the summation that goes over all infinite atoms in the nanostructure, $\L$ is the maximum number of angular momentum channels in the power spectrum, $(.)^*$ denotes the complex conjugate, $\Ne$ is the number of different chemical species (or elements) present in the system, $\Nr$ is the number of radial basis functions, and  $\cnlm$ is the product of a spherical harmonic function and a radial function with compact support, details of which can be found in \ref{appendix:cnlm}.

The atomic forces can be obtained from the energy as follows:
\begin{align}\label{Eqn:force_mlff}
\f_{\K_\alpha} &= -\frac{\partial \E}{\partial \R_{\K_\alpha}} = \sum_{\I=1}^{\N} \sum_{\t=1}^{\Nt} \wt \xi \k(\bX_{\I},\bXt_{\t}) \frac{\partial \bX_{\I}}{\partial \R_{\K_\alpha}}  \cdot \bXIt \, \forall \, \K = 1, 2, 3, \ldots,\N \,; \, \alpha = 1, 2, 3 \,,
\end{align}
where
\begin{align}\label{Eqn:XIt}
\bXIt = \frac{\bX_{\I}}{\bX_{\I} \cdot \bX_{\I}}  - \frac{\bXt_{\t}}{\bX_{\I} \cdot \bXt_{\t}} \,,
\end{align}
and the first order derivative of the atomic descriptor is given by:
\begin{align}\label{Eqn:DX}
\frac{\partial \bX_{\I}}{\partial \R_{\K_\alpha}} = \diag\left(\mathlarger{\bigoplus}_{\l=0}^\L \sqrt{\frac{8 \pi^2}{2\l+1}} \sum_{\m=-\l}^\l \left(\frac{\partial \bC_{\I,\l, \m}}{\partial \R_{\K_\alpha}}  \otimes \bC^{*}_{\I,\l,\m} + \bC_{\I,\l, \m} \otimes \frac{\partial \bC^{*}_{\I,\l,\m}}{\partial \R_{\K_\alpha}} \right) \right) \,,
\end{align} 
with
\begin{align}\label{Eqn:DC}
\frac{\partial \bC_{\I,\l, \m}}{\partial \R_{\K_\alpha}} &= \mathlarger{\mathlarger{\bigoplus}}_{\e=1}^{\Ne} \mathlarger{\mathlarger{\bigoplus}}_{\n=1}^{\Nr} \left[\sideset{}{'}\sum_{\J} \delta_{\Z_\e \Z_\J} \Big(\delta_{\bR_\K \bR_\J} - \delta_{\bR_\K \bR_\I} \Big) \cnlm^{(1)}(\bR_\J - \bR_{\I}) \right]_\alpha \,.
\end{align}
Above, $\delta_{\bR_\I \bR_\J} = \delta_{\I \J}$, $[\cdot]_\alpha$ denotes the $\alpha^{\text{th}}$ component of the vector and $\cnlm^{(1)}$ denotes the first order derivative of $\cnlm$, the expression for which is given in Eqn.~\ref{eqn:dcnlm}.

The lattice vibrations of the nanostructure are governed by the eigenproblem: 
\begin{align}\label{Eqn:phonon_eigenvalue}
\bD \bu = \omega^2 \bu \,,
\end{align}
where $\bD$  is the dynamical matrix:
\begin{align} \label{Eqn:Dyn_global}
   \left[\bD\right]_{\K_\alpha,\Q_\beta} = \frac{1}{\sqrt{\M_\K \M_\Q}} \frac{\partial^2 \E}{\partial \R_{\K_\alpha} \partial \R_{\Q_\beta}} \, \forall \, \K,\Q = 1,2,3, \ldots,\N \,; \, \alpha,\beta = 1,2,3 \,,
\end{align}
$\omega$ is the phonon frequency, and $\bu$ is the corresponding mode. Above, $\left[\bD\right]_{\K_\alpha,\Q_\beta}$ denotes the ${(\K_\alpha,\Q_\beta)}^\text{th}$ element of the dynamical matrix, $\M_\K$ is the atomic mass of the $\K^\text{th}$ atom, and the second order derivative of the energy with respect to atomic positions takes the form:
\begin{align} \label{Eqn:d2E}
\frac{\partial^2 \E}{\partial \R_{\K_\alpha} \partial \R_{\Q_\beta}} = &-\sum_{\I=1}^{\N} \sum_{\t=1}^{\Nt} \wt \xi \k(\bX_{\I},\bXt_{\t})\left( \frac{\partial}{\partial \R_{\K_\alpha}} - \xi \left(\frac{\partial \bX_{\I}}{\partial \R_{\K_\alpha}} \cdot \bXIt \right) \right) \left(\bXIt \cdot \frac{\partial \bX_{\I}}{\partial \R_{\Q_\beta}} \right) \,,
\end{align}
where
\begin{align}\label{Eqn:dXIt}
\frac{\partial \bXIt}{\partial \R_{\K_\alpha}} = \frac{1}{\bX_{\I} \cdot \bX_{\I}}\frac{\partial \bX_{\I}}{\partial \R_{\K_\alpha}}  - \frac{2\bX_{\I}}{\bX_{\I} \cdot \bX_{\I}} \left(\frac{\partial \bX_{\I}}{\partial \R_{\K_\alpha}} \cdot \frac{\bX_\I}{\bX_{\I} \cdot \bX_{\I}}\right)
 + \frac{\bXt_{\t}}{\bX_{\I} \cdot \bXt_{\t}} \left(\frac{\partial \bX_{\I}}{\partial \R_{\K_\alpha}} \cdot \frac{\bXt_\t}{\bX_{\I} \cdot \bXt_{\t}}\right) \,.
\end{align}
In the above expression, the second order derivative of the atomic descriptor is given by:
\begin{align}
\frac{\partial^2 \bX_{\I}}{\partial \R_{\K_\alpha} \partial \R_{\Q_\beta}} = &\diag\Bigg(\mathlarger{\bigoplus}_{\l=0}^\L \sqrt{\frac{8 \pi^2}{2\l+1}} \sum_{\m=-\l}^\l \bigg( \frac{\partial^2 \bC_{\I,\l, \m}}{\partial \R_{\K_\alpha} \partial \R_{\Q_\beta}}  \otimes \bC^{*}_{\I,\l,\m} + \frac{\partial \bC_{\I,\l, \m}}{\partial \R_{\K_\alpha}}  \otimes \frac{\partial \bC^*_{\I,\l,\m}}{\partial \R_{\Q_\beta}} \nonumber \\
&+ \frac{\partial \bC_{\I,\l, \m}}{\partial \R_{\Q_\beta}}  \otimes \frac{\partial \bC^*_{\I,\l,\m}}{\partial \R_{\K_\alpha}} + \bC_{\I,\l, \m} \otimes \frac{\partial^2 \bC^{*}_{\I,\l,\m}}{\partial \R_{\K_\alpha} \partial \R_{\Q_\beta}} \bigg) \Bigg) \,,
\end{align}
where
\begin{align} \label{Eqn:DC2}
\frac{\partial^2 \bC_{\I,\l, \m}}{\partial \R_{\K_\alpha}\partial \R_{\Q_\beta}} &= \mathlarger{\mathlarger{\bigoplus}}_{\e=1}^{\Ne} \mathlarger{\mathlarger{\bigoplus}}_{\n=1}^{\Nr} \left[\sideset{}{'} \sum_{\J} \delta_{\Z_\e \Z_\J} \left( \delta_{\bR_\Q \bR_\J} -\delta_{\bR_\Q \bR_\I} \right) \left(\delta_{\bR_\K \bR_\J} - \delta_{\bR_\K \bR_\I} \right)\cnlm^{(2)} (\bR_\J - \bR_{\I})\right]_{\alpha \beta} \,.
\end{align}
Above, $\cnlm^{(2)}$ is a $3\times3$ matrix denoting the second order derivative of $\cnlm$, whose expression can be found in Eqn.~\ref{eqn:d2cnlm}. 
%%%%%%%%%%%%%%%%%%%%%%%%%%%%%%%%%%%%%%%%%%%%%%%%%%%%%%%%%%%%%%%%%%%%%%%%%
\subsection{Symmetry-adapted SOAP model} \label{Subsec:SOAP:Symmetry}

We now exploit the cyclic and helical symmetry in the nanostructure to develop a  symmetry-adapted version of the SOAP model. In particular, we first derive the relationships obeyed by the atomic descriptors and their derivatives, for both the fundamental atoms and their images. As an initial step, we derive these relationships for $\C$ and its derivatives, where $\C$ is a diagonal element of the $\bC_{\I,\l, \m}$ matrix. Henceforth, we will  use $\bR_\I^{\zeta,\mu}$ to denote $\sfgm \circ \bR_\I$, where $\I \in [1,\hN]$ and  $\sfgm \in \sfGp$, unless otherwise stated.

The relationship between $\C$ corresponding to different images of a fundamental atom can be derived as:
\begin{align} \label{Eqn:C_transform}
\C(\bR_{\I}^{\zeta, \mu}) &= \sideset{}{'} \sum_{\sfgmb \in \sfG} \sideset{}{'} \sum_{\J=1}^{\hN} \delta_{\Z_\e \Z_\J} \cnlm(\bR_\J^{\bar{\zeta},\bar{\mu}} - \bR_{\I}^{\zeta,\mu}) \nonumber \\
& = \mathrm{e}^{-i \m (\zeta \Theta + \mu \varphi)} \sideset{}{'} \sum_{\sfgmb \in \sfG} \sideset{}{'} \sum_{\J=1}^{\hN} \delta_{\Z_\e \Z_\J} \cnlm(\bR_\J^{\bar{\zeta}-\zeta,\bar{\mu}-\mu} - \bR_{\I}) \nonumber \\
& = \mathrm{e}^{-i \m (\zeta \Theta + \mu \varphi)} \C(\bR_\I) \,,
\end{align}
where the second equality comes from the fact that $\sfG$ is a set of isometries and that the spherical harmonic function transforms with a phase factor under the action of an element of $\sfG$ (\ref{appendix:sphericalfunc_transform}), and the last equality follows from the closure property of $\sfG$. Similarly, the relationship between the first order derivative of $\C$ corresponding to different images of a fundamental atom can be derived as:
\begin{align}\label{Eqn:DC_transform}
\frac{\partial \C (\bR_{\I}^{\zeta,\mu})}{\partial \bR_{\K}^{\hat{\zeta},\hat{\mu}}} & = \sideset{}{'} \sum_{\sfgmb \in \sfG}\sideset{}{'} \sum_{\J=1}^{\hN} \delta_{\Z_\e \Z_\J} \Big(\delta_{\bR_{\K}^{\hat{\zeta},\hat{\mu}} \bR_\J^{\bar{\zeta},\bar{\mu}}} - \delta_{\bR_{\K}^{\hat{\zeta},\hat{\mu}} \bR_{\I}^{\zeta,\mu}}\Big) \cnlm^{(1)}(\bR_\J^{\bar{\zeta},\bar{\mu}} - \bR_{\I}^{\zeta,\mu}) \nonumber \\
&= \mathrm{e}^{-i m (\zeta \Theta + \mu \varphi)} \rotM \sideset{}{'} \sum_{\sfgmb \in \sfG}\sideset{}{'}\sum_{\J=1}^{\hN} \delta_{\Z_\e \Z_\J} \Big(\delta_{\bR_{\K}^{\hat{\zeta},\hat{\mu}} \bR_\J^{\bar{\zeta},\bar{\mu}}} - \delta_{\bR_{\K}^{\hat{\zeta},\hat{\mu}} \bR_{\I}^{\zeta,\mu}}\Big) \cnlm^{(1)}(\bR_\J^{\bar{\zeta}-\zeta,\bar{\mu}-\mu} - \bR_{\I})  \nonumber \\
&= \mathrm{e}^{-i m (\zeta \Theta + \mu \varphi)} \rotM \sideset{}{'} \sum_{\sfgmb \in \sfG}\sideset{}{'}\sum_{\J=1}^{\hN} \delta_{\Z_\e \Z_\J} \Big(\delta_{\bR_{\K}^{\hat{\zeta},\hat{\mu}} \bR_\J^{\bar{\zeta}+\zeta,\bar{\mu}+\mu}} - \delta_{\bR_{\K}^{\hat{\zeta},\hat{\mu}} \bR_{\I}^{\zeta,\mu}}\Big) \cnlm^{(1)}(\bR_\J^{\bar{\zeta},\bar{\mu}} - \bR_{\I})  \nonumber \\
&= \mathrm{e}^{-i m (\zeta \Theta + \mu \varphi)} \rotM \sideset{}{'} \sum_{\sfgmb \in \sfG}\sideset{}{'}\sum_{\J=1}^{\hN} \delta_{\Z_\e \Z_\J} \Big(\delta_{\bR_{\K}^{\hat{\zeta}-\zeta,\hat{\mu}-\mu} \bR_\J^{\bar{\zeta},\bar{\mu}}} - \delta_{\bR_{\K}^{\hat{\zeta}-\zeta,\hat{\mu}-\mu} \bR_{\I}}\Big) \cnlm^{(1)}(\bR_\J^{\bar{\zeta},\bar{\mu}} - \bR_{\I})  \nonumber \\
&= \mathrm{e}^{-i m (\zeta \Theta + \mu \varphi)} \rotM \frac{\partial \C (\bR_\I)}{\partial \bR_{\K}^{\hat{\zeta}-\zeta,\hat{\mu}-\mu}} \,,
\end{align} 
where $\delta_{\bR_{\I}^{\zeta,\mu} \bR_\J^{\bar{\zeta},\bar{\mu}}} = \delta_{\I \J} \delta_{\zeta \bar{\zeta}} \delta_{\mu \bar{\mu}}$ and $\sfgmh \in \sfG$. Above, the second equality follows from the transformation of the first order derivative of the radial and spherical harmonic functions under the action of an element of $\sfG$ (\ref{appendix:sphericalfunc_transform}), the third equality is obtained by replacing $(\bar{\zeta},\bar{\mu})$ with $(\bar{\zeta} + \zeta, \bar{\mu} + \mu)$ and using the closure property of $\sfG$, and the fourth equality comes from the symmetry of the nanostructure. Lastly, the relationship between the second order derivative of $\C$ corresponding to different images of a fundamental atom can be derived as::
\begin{align} \label{Eqn:D2C_transform}
\frac{\partial^2 \C (\bR_{\I}^{\zeta,\mu})}{\partial \bR_\K^{\hat{\zeta},\hat{\mu}}\partial \bR_\Q^{\tilde{\zeta},\tilde{\mu}}} &= \sideset{}{'} \sum_{\sfgmb \in \sfG}\sideset{}{'} \sum_{\J=1}^{\hN} \delta_{\Z_\e \Z_\J} \Big(\delta_{\bR_{\Q}^{\tilde{\zeta},\tilde{\mu}} \bR_\J^{\bar{\zeta},\bar{\mu}}} - \delta_{\bR_{\Q}^{\tilde{\zeta},\tilde{\mu}} \bR_{\I}^{\zeta,\mu}}\Big) \Big(\delta_{\bR_{\K}^{\hat{\zeta},\hat{\mu}} \bR_\J^{\bar{\zeta},\bar{\mu}}} - \delta_{\bR_{\K}^{\hat{\zeta},\hat{\mu}} \bR_{\I}^{\zeta,\mu}}\Big) \cnlm^{(2)} (\bR_\J^{\bar{\zeta},\bar{\mu}} - \bR_{\I}^{\zeta,\mu}) \nonumber \\
&= \mathrm{e}^{-i m (\zeta \Theta + \mu \varphi)} \rotM \Bigg(\sideset{}{'} \sum_{\sfgmb \in \sfG}\sideset{}{'} \sum_{\J=1}^{\hN} \delta_{\Z_\e \Z_\J} \Big(\delta_{\bR_{\Q}^{\tilde{\zeta},\tilde{\mu}} \bR_\J^{\bar{\zeta},\bar{\mu}}} - \delta_{\bR_{\Q}^{\tilde{\zeta},\tilde{\mu}} \bR_{\I}^{\zeta,\mu}}\Big) \Big(\delta_{\bR_{\K}^{\hat{\zeta},\hat{\mu}} \bR_\J^{\bar{\zeta},\bar{\mu}}} - \delta_{\bR_{\K}^{\hat{\zeta},\hat{\mu}} \bR_{\I}^{\zeta,\mu}}\Big) \nonumber \\
& \quad\times \cnlm^{(2)} (\bR_\J^{\bar{\zeta}-\zeta,\bar{\mu}-\mu} - \bR_{\I}) \Bigg)\rotMinv \nonumber \\
&= \mathrm{e}^{-i m (\zeta \Theta + \mu \varphi)} \rotM \Bigg(\sideset{}{'} \sum_{\sfgmb \in \sfG}\sideset{}{'} \sum_{\J=1}^{\hN} \delta_{\Z_\e \Z_\J} \Big(\delta_{\bR_{\Q}^{\tilde{\zeta}-\zeta,\tilde{\mu} -\mu} \bR_\J^{\bar{\zeta},\bar{\mu}}} - \delta_{\bR_{\Q}^{\tilde{\zeta}-\zeta,\tilde{\mu}-\mu} \bR_{\I}}\Big) \Big(\delta_{\bR_{\K}^{\hat{\zeta}-\zeta,\hat{\mu}-\mu} \bR_\J^{\bar{\zeta},\bar{\mu}}} \nonumber \\
&\quad - \delta_{\bR_{\K}^{\hat{\zeta}-\zeta,\hat{\mu}-\mu} \bR_{\I}}\Big) \cnlm^{(2)} (\bR_\J^{\bar{\zeta},\bar{\mu}} - \bR_{\I}) \Bigg)\rotMinv \nonumber \\
&= \mathrm{e}^{-i m (\zeta \Theta + \mu \varphi)} \rotM \frac{\partial^2 \C (\bR_{\I})}{\partial \bR_\K^{\hat{\zeta}-\zeta,\hat{\mu}-\mu}\partial \bR_\Q^{\tilde{\zeta}-\zeta,\tilde{\mu}-\mu}} \rotMinv \,,
\end{align}
where $\sfgmh \in \sfG, \sfgmt \in \sfG$. Above, the second equality  follows from the transformation of the second order derivative of the radial and spherical harmonic functions under the action of an element of $\sfG$ (\ref{appendix:sphericalfunc_transform}), and the third equality is obtained by replacing $(\bar{\zeta},\bar{\mu})$ with $(\bar{\zeta} + \zeta, \bar{\mu} + \mu)$ and using the closure property of $\sfG$.

Utilizing Eqns.~\eqref{Eqn:C_transform}---\eqref{Eqn:D2C_transform}, we arrive at the relationships satisfied by the atomic descriptor and its derivatives, corresponding to different images of a  fundamental atom:
\begin{align}
\bX_{\I}^{\zeta,\mu} &= \bX_{\I} \,,\label{Eqn:Desc_transform}\\
\frac{\partial \bX_{\I}^{\zeta,\mu}}{\partial \bR_\K^{\hat{\zeta},\hat{\mu}}} &= \frac{\partial \bX_{\I}}{\partial \bR_\K^{\hat{\zeta}-\zeta,\hat{\mu}-\mu}} \rotMinv\,, \label{Eqn:dDesc_transform}\\
\frac{\partial^2 \bX_{\I}^{\zeta,\mu}}{\partial \bR_\K^{\hat{\zeta},\hat{\mu}}\partial \bR_\Q^{\tilde{\zeta},\tilde{\mu}}} &= \rotM \frac{\partial}{\bR_\K^{\hat{\zeta}-\zeta}} \left( \frac{\partial \bX_{\I}}{\partial \bR_\Q^{\tilde{\zeta}-\zeta,\tilde{\mu}-\mu}} \rotMinv \right) \label{Eqn:d2Desc_transform}\,,
\end{align} 
 where $\bX_\I^{\zeta,\mu}$ is the atomic descriptor corresponding to atom with atomic position $\bR_\I^{\zeta,\mu}$.
  
\paragraph{\underline{Symmetry-adapted energy}} It follow from Eqns.~\ref{Eqn:Energy_mlff_total} and~\ref{Eqn:Desc_transform} that the atomic energy of an image atom is identical to that of the  corresponding fundamental atom, i.e., 
\begin{align}
\E_\I^{\zeta,\mu} = \E_\I \,.
\end{align}
Thereafter, the total energy associated with the fundamental atoms, also known as fundamental domain energy, can be obtained as:
\begin{align} \label{Eqn:Energy_mlff_symadapt}
\hat{\E} = \frac{\hN}{\N} \sum_{\sfgm \in \sfGp} \sum_{\I=1}^{\hN} \E_{\I}^{\zeta,\mu} = \sum_{\I=1}^{\hN} \E_\I = \sum_{\I=1}^{\hN} \sum_{\t=1}^{\Nt} \wt \k(\bX_{\I},\bXt_{\t}) \,.
\end{align}
\paragraph{\underline{Symmetry-adapted atomic forces}}
It follow from Eqns.~\ref{Eqn:force_mlff},~\ref{Eqn:Desc_transform}, and~\ref{Eqn:dDesc_transform} that
\begin{align}
\frac{\partial \E_{\I}^{\zeta,\mu}}{\partial \bR_{\K}} =  \rotM \frac{\partial \E_{\I}}{\partial \bR_{\K}^{-\zeta,-\mu}} \,.
\end{align}
Therefore, the  atomic force for a fundamental atom can be written as:
\begin{align}\label{Eqn:force_mlff_symadapt}
\f_{\K_\alpha} &= - \sum_{\sfgm \in \sfGp} \sum_{\I=1}^{\hN} \frac{\partial \E_{\I}^{\zeta,\mu}}{\partial \R_{\K_\alpha}} = - \sum_{\sfgm \in \sfGp} \sum_{\I=1}^{\hN} \left[\rotM \frac{\partial \E_{\I}}{\partial \bR_{\K}^{-\zeta,-\mu}}\right]_\alpha \nonumber \\
&= \sum_{\sfgm \in \sfGp} \sum_{\I=1}^{\hN} \sum_{\t=1}^{\Nt} \wt \xi \k(\bX_{\I},\bXt_{\t}) \left[ \left(\frac{\partial \bX_{\I}}{\partial \bR_{\K}^{-\zeta,-\mu}} \rotMinv \right) \cdot \bXIt \right]_\alpha \,.
\end{align}
In the absence of symmetry-breaking non-isomorphic\footnote{Rigid body motion is a typical example of isometric-isomorphism. External twist on nanotubes is an example of homeomorphic-isomorphism. These isomorphisms are symmetry-preserving and hence can be studied using the developed framework.} external disturbance, the atomic forces on the nanostructure must satisfy the relation:
\begin{align}
\bf_{\I}^{\hat{\zeta},\hat{\mu}} = \rotMh \bf_\I \quad \forall \, \I \in [1,\hN], \sfgmh \in \sfGp \,,
\end{align}
which translates to the following relation within the symmetry-adapted SOAP framework:
\begin{align}
&\sum_{\sfgm \in \sfGp}  \frac{\partial \bX_{\I}}{\partial \bR_{\I}^{\hat{\zeta}-\zeta,\hat{\mu}-\mu}} \rotMinv = \sum_{\sfgm \in \sfGp}  \frac{\partial \bX_{\I}}{\partial \bR_{\I}^{-\zeta,-\mu}} \rotMMhinv \nonumber \\
\implies & \sum_{\substack{\sfgmt - \sfgmh \\ \sfgmt \in \sfGp}} \frac{\partial \bX_{\I}}{\partial \bR_{\I}^{-\tilde{\zeta},-\tilde{\mu}}} \rotMMtinv = \sum_{\sfgm \in \sfGp}  \frac{\partial \bX_{\I}}{\partial \bR_{\I}^{-\zeta,-\mu}} \rotMMhinv \,.
\end{align}
This relation is  true if and only if $\sfGp = \sfG$. However, since the descriptors have a compact support, there may only be a subset of atoms in the helical direction for which the atomic forces satisfy the above relation. Indeed, the above relation is a necessary requirement to block-diagonalize the dynamical matrix, from which symmetry-adapted phonons can be obtained, as discussed next. For the current purposes, we assume the above relationship holds throughout the nanostructure. This is justified for a large enough nanostructure in which the effect of the edge atoms is negligible.

\paragraph{\underline{Symmetry-adapted phonons}} Starting from the dynamical matrix organized as the orbit of the fundamental atoms under the action of $\sfGp$, the symmetry-adapted phonon eigenvalue equation is given by \citep{aghaei2013symmetry, sharmathesis}:
\begin{align}\label{Eqn:phononeig_symadapt}
\bD_\bq \bu_\bq = \omega^2_\bq \bu_\bq \,,
\end{align}
where $\bD_\bq$ is the symmetry-adapted dynamical matrix whose ${(\K_\alpha,\Q_\beta)}^\text{th}$ element is given by:
\begin{align}
   \left[\bD_\bq\right]_{\K_\alpha,\Q_\beta} = \frac{1}{\sqrt{\M_\K \M_\Q}} \frac{\hN}{\N} \sum_{\sfgmh \in \sfGp} \sum_{\sfgmt \in \sfGp} \mathrm{e}^{i \left( \nu_\bq(\tilde{\zeta} - \hat{\zeta}) \Theta + \eta_\bq (\tilde{\mu}-\hat{\mu}) \tau \right)} \left[ \rotMhinv \frac{\partial^2 \E}{\partial {\bR}_\K^{\hat{\zeta},\hat{\mu}} \partial \bR_\Q^{\tilde{\zeta},\tilde{\mu}}} \rotMt \right]_{\alpha \beta} \nonumber \\
   \, \forall \, \K,\Q = 1,2,3,\ldots,\hN \,; \, \alpha,\beta = 1,2,3 \,.
\end{align}
Above, $\omega_\bq$ and $\bu_\bq$ are the phonon frequency and mode, respectively,  corresponding to the phonon wavevector $\bq = (0,\nu_\bq,\eta_\bq)$, where $\nu_\bq \in \{0,1,2, \ldots, \mathfrak{N}-1\}$ and $\eta_\bq \in [-\frac{\pi}{\tau},\frac{\pi}{\tau})$. It follows from Eqns.~\ref{Eqn:d2E},~\ref{Eqn:C_transform},~\ref{Eqn:DC_transform}, and~\ref{Eqn:D2C_transform} that:
\begin{align}
\frac{\partial^2 \E_{\I}^{\zeta,\mu}}{\partial \bR_\K \partial \bR_\Q} = \rotM \frac{\partial^2 \E_{\I}}{\partial \bR_\K^{-\zeta,-\mu} \partial \bR_\Q^{-\zeta,-\mu}} \rotMinv\,.
\end{align}
Using the above relationship, the symmetry-adapted dynamical matrix can be further reduced as:
\begin{align} \label{Eqn:Dyn_local}
   \left[\bD_\bq\right]_{\K_\alpha,\Q_\beta} =& \frac{1}{\sqrt{\M_\K \M_\Q}} \frac{\hN}{\N} \sum_{\sfgmh \in \sfGp} \sum_{\sfgmt  \in \sfGp} \sum_{\sfgm  \in \sfGp} \sum_{\I=1}^{\hN} \mathrm{e}^{i \left( \nu_\bq(\tilde{\zeta} - \hat{\zeta}) \Theta + \eta_\bq (\tilde{\mu}-\hat{\mu}) \tau \right)} \left[ \rotMhinv \frac{\partial^2 \E_\I^{\zeta,\mu}}{\partial \bR_\K^{\hat{\zeta},\hat{\mu}} \partial \bR_\Q^{\tilde{\zeta},\tilde{\mu}}} \rotMt \right]_{\alpha \beta} \nonumber \\
   =& \frac{1}{\sqrt{\M_\K \M_\Q}}  \sum_{\I=1}^{\hN} \sum_{\sfgmh  \in \sfGp} \sum_{\sfgmt  \in \sfGp} \mathrm{e}^{i \left( \nu_\bq(\tilde{\zeta} - \hat{\zeta}) \Theta + \eta_\bq (\tilde{\mu}-\hat{\mu}) \tau \right)} \left[ \rotMhinv \frac{\partial^2 \E_\I}{\partial \bR_\K^{\hat{\zeta},\hat{\mu}} \partial \bR_\Q^{\tilde{\zeta},\tilde{\mu}}} \rotMt \right]_{\alpha \beta} \nonumber \\
   =&- \frac{1}{\sqrt{\M_\K \M_\Q}}  \sum_{\I=1}^{\hN} \sum_{\t=1}^{\Nt} \wt \xi \k(\bX_{\I},\bXt_{\t}) \sum_{\sfgmh  \in \sfGp} \sum_{\sfgmt  \in \sfGp} \mathrm{e}^{i \left( \nu_\bq(\tilde{\zeta} - \hat{\zeta}) \Theta + \eta_\bq (\tilde{\mu}-\hat{\mu}) \tau \right)} \Bigg[ \bigg(\rotMhinv \frac{\partial}{\partial \bR_{\K}^{\hat{\zeta},\hat{\mu}}} \nonumber \\
   & - \xi \left(\frac{\partial \bX_{\I}}{\partial \bR_{\K}^{\hat{\zeta},\hat{\mu}} }\rotMh\right) \cdot \bXIt \bigg) \bigg(\bXIt \cdot \frac{\partial \bX_{\I}}{\partial \bR_{\Q}^{\tilde{\zeta},\tilde{\mu}}} \rotMt\bigg) \Bigg]_{\alpha \beta}\,.
\end{align}

It is worth noting that by suitably adapting the rotational matrices and the phase factors in the above-derived symmetry-adapted expressions, the above formalism is applicable to other symmetries found in 1D nanostructures. For example, in the case of cyclic and translational symmetry, the rotational matrix corresponding to helical symmetry in the above expressions will be replaced by a $3\times3$ identity matrix, with $\eta_\bq$ component of the wavevector present in the Brillouin zone of the corresponding 1D periodic cell.

%%%%%%%%%%%%%%%%%%%%%%%%%%%%%%%%%%%%%%%%%%%%%%%%%%%%%%%%%%%%%%%%%%%%%%%%%
\subsection{Model weights} \label{Subsec:Training}
Given the energy and atomic forces corresponding to training nanostructures (also known as reference nanostructures) and a set of training descriptors, the model weights can be determined from the following minimization problem \citep{kumar2024fly}:
\begin{align} \label{Eq:Optw}
\bw = \arg \min_{\bw} \left[ \frac{\beta}{2}  \left( \wE^2 \Big| \yE-\KE\bw \Big|^2 + \wf^2\Big| \yf- \frac{\sigE}{\sigf}\Kf\bw \Big|^2\right) + \frac{\alpha}{2} |\bw|^2 \right] \,,
\end{align}
where $\bw \in \mathbb{R}^{\Nt}$ is the vector of model weights; $\alpha$ and $\beta$ are the Bayesian parameters; $\wE$ and $\wf$ are the weighting factors for the errors in energy and forces, respectively; $\yE \in \mathbb{R}^{\Nst}$ and $\yf \in \mathbb{R}^{3 \sum_{A=1}^{\Nst} \hN}$ are the scaled and shifted vectors of energy and forces corresponding to $\Nst$ training nanostructures, respectively; $\sigE$ and $\sigf$ are the standard deviations in energy and forces across all training nanostructures, respectively; $\KE \in \mathbb{R}^{\Nst \times \Nt}$ and $\Kf \in \mathbb{R}^{3 \sum_{A=1}^{\Nst} \hN \times \Nt}$ are the covariance matrices:
\begin{align}
[\KE]_{A,\t} &= \sum_{\I=1}^{\hN} \k({\bX_A}_{\I},\bXt_\t)\,, \\
[\Kf]_{A_{\K_\alpha},\t} &= \sum_{\sfgm \in \sfGp_A} \sum_{\I=1}^{\hN} \xi \k({\bX_A}_{\I},\bXt_\t) \left[ \left(\frac{\partial {\bX_A}_{\I}}{\partial \bR_{A_\K}^{-\zeta,-\mu}} \mathbf{Q}^{-\zeta}_{\Theta_A} \mathbf{Q}^{-\mu}_{\varphi_A} \right) \cdot \mathbf{W}_{A_a,\t} \right]_\alpha \,.
\end{align}
Above, $[\KE]_{A,\t}$ denotes the $(A,\t)^\text{th}$ element of $\KE$, $[\Kf]_{A_{\K_\alpha},\t}$ denotes the $(3 \sum_{a=1}^{A-1} \hat{N}_a + 3*(\K-1) + \alpha,\t)^\text{th}$ element of $\Kf$, and the subscript $A$ below a variable indicates the structure to which it corresponds. The solution of the minimization problem in Eqn.~\ref{Eq:Optw} can be written as:
\begin{align} \label{Eq:optimumweights}
\bw = \beta \left(\alpha \bI +\beta\left(\KEt^T \KEt + \Kft^T \Kft\right) \right)^{-1}\left(\KEt^T \yEt +\Kft^T \yft \right)\,,
\end{align}
where $\KEt = \wE \KE$, $\Kft = \wf \frac{\sigE}{\sigf} \Kf$, $\yEt = \wE \yE$, and $\yft = \wf \yf$. Once the model weights are computed, the energy, forces, and phonons for other nanostructures can be predicted. 
%%%%%%%%%%%%%%%%%%%%%%%%%%%%%%%%%%%%%%%%%%%%%%%%%%%%%%%%%%%%%%%%%%%%%%%%%

%%%%%%%%%%%%%%%%%%%%%%%%%%%%%%%%%%%%%%%%%%%%%%%%%%%%%%%%%%%%%%%%%%%%%%%%%%%%%%%%%%%%%%%%%%%%%%%
\section{Results and discussion} \label{Sec:Results}

\begin{figure}[htbp!]
    \centering
\includegraphics[keepaspectratio=true,width=0.75\textwidth]{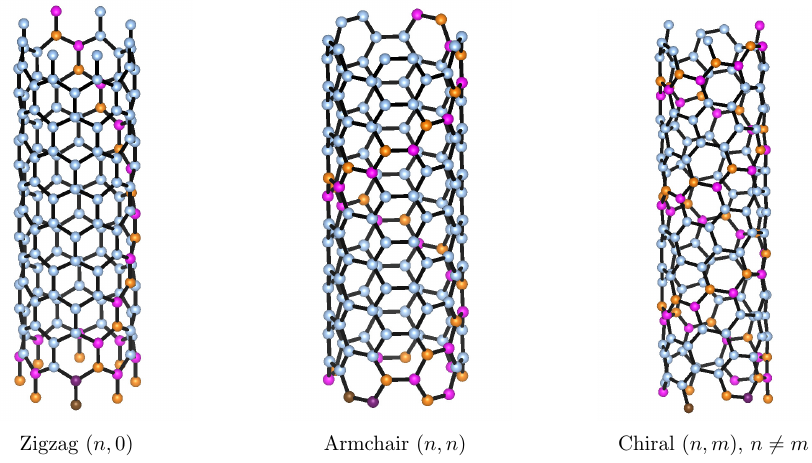}    
    \caption{Illustration of single-walled CNTs. The two fundamental carbon atoms are highlighted in dark yellow and magenta, with their cyclic and helical images having lighter shades.}
    \label{Fig:CNT_chirality}
\end{figure}

We have implemented the  formulation for cyclic and helical symmetry-informed MLFFs  in the SPARC electronic structure code \citep{xu2021sparc, zhang2024sparc}. We now apply this framework to study lattice vibrations in single-walled carbon nanotubes (CNTs). A CNT is a cylindrical nanostructure generated by the action of a cyclic and helical symmetry group on the two fundamental carbon atoms in monolayer graphene. Based on the symmetry groups and the fundamental atoms chosen, carbon nanotubes can be classified into three main categories, namely, i) zigzag $(n,0)$,  iii) armchair $(n,n)$, and iii) chiral $(n,m)$, $n \neq m$, as illustrated in Fig.~\ref{Fig:CNT_chirality}. The $(n,m)$ representation of the nanotube is indicative of  its chirality, with zigzag and armchair nanotubes referred to as achiral, the remaining referred to as chiral. The unrelaxed radius of the CNT can also be determined from its chirality using the relation: $r = \frac{a0}{2 \pi}\sqrt{3(n^2+m^2+nm)}$, where $a0$ is the carbon-carbon bond length in graphene. Below, we first develop a symmetry-informed MLFF for CNTs in Section~\ref{Sec:MLFFforCNT}, and then apply it to study phonons in CNTs in Section~\ref{Sec:phononsinCNT}. 

%%%%%%%%%%%%%%%%%%%%%%%%%%%%%%%%%%%%%%%%%%%%%%%%%%%%%%%%%%%%%%%%%%%%%%%%%
\subsection{Symmetry-informed MLFF for carbon nanotubes} \label{Sec:MLFFforCNT}

%%%%%%%%%%%%%%%%%%%%%%%%%%
\paragraph{\underline{Training}}  
Choosing a representative set of 7 different carbon nanotubes with varying diameter and chirality, namely, $(20,0)$, $(24,0)$, $(28,0)$, $(10,10)$, $(16,16)$, $(12,6)$, and $(18,9)$, we train the symmetry-informed MLFF model by  performing on-the-fly \citep{jinnouchi2019fly,kumar2023kohn} symmetry-adapted MLFF molecular dynamics (MD) simulation for each of these nanotubes. In particular, we perform the DFT calculations using the Cyclix-DFT feature of SPARC \citep{sharma2021real}, while employing the local density approximation (LDA) \citep{Kohn1965} for the exchange-correlation functional, and optimized norm-conserving Vanderbilt (ONCV) \citep{hamann2013optimized} pseudopotential for carbon from the SPMS table \citep{spms}. Cyclix-DFT is now a mature feature that has been used in a range of applications, including the twisting of 1D nanostructures \citep{bhardwaj2022strain, bhardwaj2021torsional, bhardwaj2022elastic, bhardwaj2021torsional, bhardwaj2023ab, bhardwaj2024strain} and bending of 2D materials \citep{kumar2020bending, kumar2021flexoelectricity, kumar2022bending, codony2021transversal}. All parameters in the Cyclix-DFT calculations are chosen such that the energies and forces are converged to within $10^{-5}$ Ha/atom and $10^{-4}$ Ha/Bohr, respectively. The MLFF model parameters are chosen to be the same as previous work \citep{kumar2024fly, kumar2023kohn, kumar2024shock}: 6 angular momentum channels, 8 radial basis functions, and $10$ Bohr cutoff radius for atomic descriptors. The MD simulations are performed at room temperature with a time step of 1 fs, using the microcanonical (NVE) ensemble that is driven by velocity Verlet integration of Newton's equations of motion \citep{frenkel2023understanding}. The initial atomic configuation for the MD is generated by first performing the symmetry-adapted geometry relaxation on the fundamental atoms, followed by the application of the  cyclic and helical symmetry operator  to generate a larger system of $\sim$ 50 atoms that is used for the MD. This process results in a computationally efficient sampling of diverse and physically relevant atomic environments near the relaxed atomic configuration of each nanotube, yielding a comprehensive set of training descriptors. Moreover, it also generates training data consisting of DFT energies and atomic forces corresponding to only those atomic configurations for which the model prediction error exceeds the threshold. Finally, we perform CUR downsampling \citep{young2021transferable,jinnouchi2019fly} on the generated descriptors to obtain a linearly independent set of training descriptors. Once the training descriptors and the training data corresponding to  all the aforementioned training nanotubes are generated, we compute the model weights, the resulting model henceforth referred to as CNT-MLFF.  In the current work, the total number of carbon nanostructures and CUR sparsified descriptors employed in training CNT-MLFF are 182 and 449, respectively.
%%%%%%%%%%%%%%%%%%%%%%%%%%%%%%%%%%%%%

\paragraph{\underline{Accuracy}}

\begin{figure}[htbp!]
    \centering
    \begin{subfigure}[b]{0.49\textwidth}
        \centering
        \includegraphics[keepaspectratio=true,width=\textwidth]{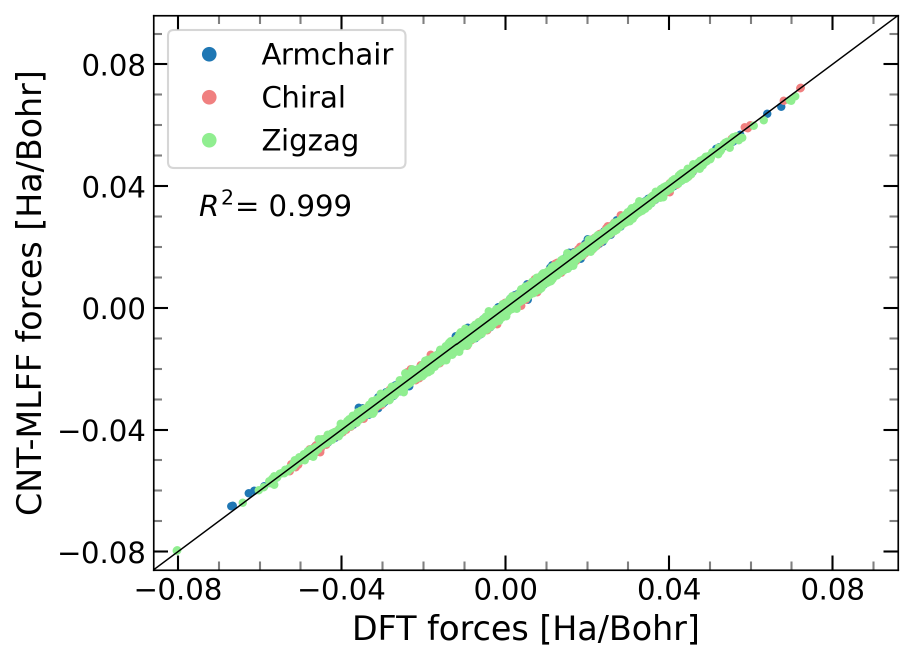}
        \caption{Training error}
        \label{fig:dF_training}
    \end{subfigure} %\hfill
    \begin{subfigure}[b]{0.49\textwidth}
        \centering
        \includegraphics[keepaspectratio=true,width=\textwidth]{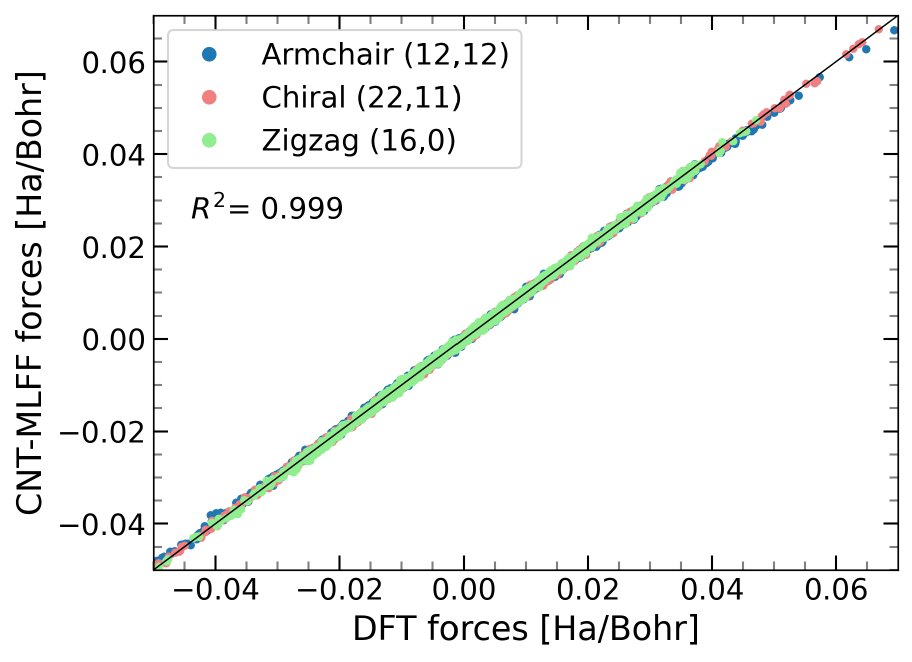}
        \caption{Prediction error}
        \label{fig:dF_validation}
    \end{subfigure}
    \caption{Training and prediction errors in the forces for the CNT-MLFF model.}
    \label{Fig:Forces error}
\end{figure}

First, we verify the quality of the fitting in CNT-MLFF by computing the training error in the energy and atomic forces. The results so obtained for the forces are presented in Fig.~\ref{fig:dF_training}, where we plot the deviation of the CNT-MLFF forces from those computed by DFT for the atomic configurations encountered during the training steps.  It is clear from the results that the machine learned model is able to accurately fit the DFT data. In particular, the mean absolute error (MAE) and root mean square error (RMSE) in energy are $5.6 \times 10^{-5}$ Ha/atom and $6.5 \times 10^{-5}$ Ha/atom, respectively, while the corresponding numbers for the forces are $3.3 \times 10^{-4}$ Ha/Bohr and $4.6 \times 10^{-4}$ Ha/Bohr, respectively.

Next, we verify the accuracy of CNT-MLFF by computing the prediction error in the energy and atomic forces for 82 randomly chosen atomic configurations encountered during the on-the-fly MLFF MD simulations for the (12,12), (22,11), and (16,0) nanotubes. Note that these nanotubes were not part of the training for the machine learned model. We present the results so obtained for the forces in Fig.~\ref{fig:dF_validation}, where we plot the deviation of the CNT-MLFF forces from those computed by Cyclix-DFT.  All parameters in the Cyclix-DFT calculations are chosen such that the energies and forces are converged to within $10^{-5}$ Ha/atom and $10^{-4}$ Ha/bohr, respectively. It is clear from the results that the machine learned model is able to make accurate predictions for these nanotubes, even though they were not part of the training. In particular, the MAE and RMSE in energy are $1.0 \times 10^{-4}$ Ha/atom and $1.4 \times 10^{-4}$ Ha/atom, respectively, while the corresponding numbers for the forces are $4.5 \times 10^{-4}$ Ha/Bohr and $4.7 \times 10^{-4}$ Ha/Bohr, respectively. 

\begin{figure}[htbp!]
    \centering
    \begin{subfigure}[b]{0.49\textwidth}
        \centering
        \includegraphics[keepaspectratio=true,width=\textwidth]{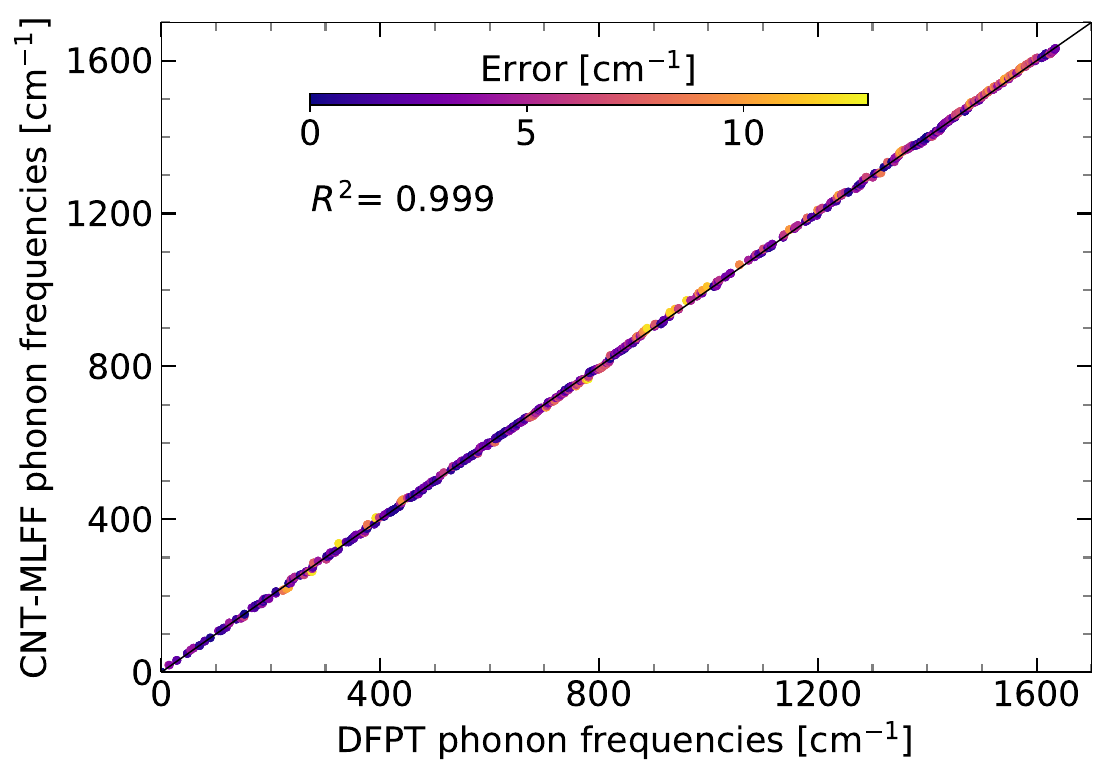}
        \caption{Phonon frequencies}
        \label{fig:dph_testing}
    \end{subfigure}
    \begin{subfigure}[b]{0.49\textwidth}
        \centering
        \includegraphics[keepaspectratio=true,width=0.955\textwidth]{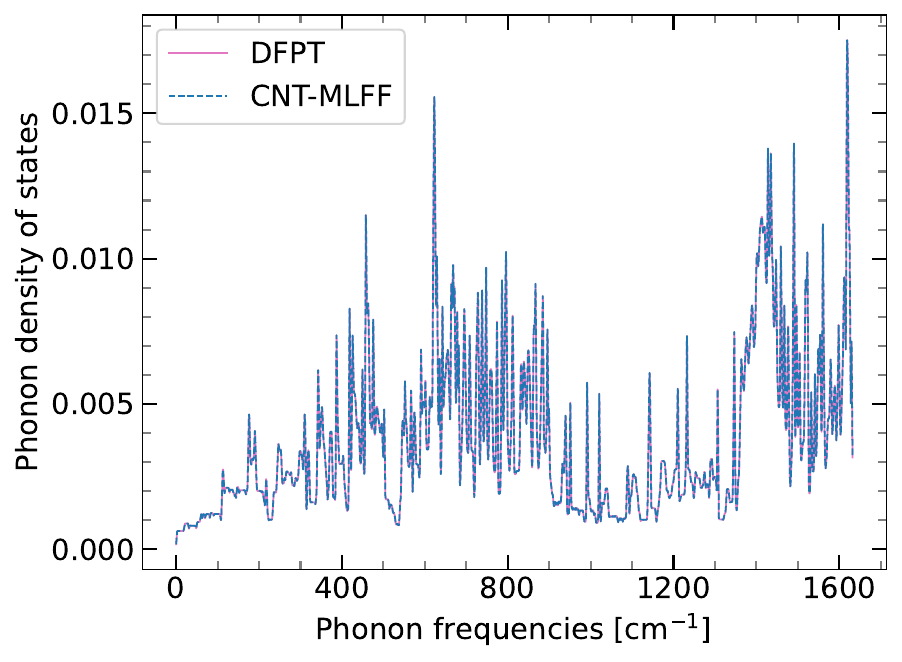}
        \caption{Phonon DOS}
        \label{fig:ph_DOS_abinit_mlff}
    \end{subfigure}
    \caption{Errors in the phonon frequencies and  DOS predicted by the CNT-MLFF model for the $(16,0)$ CNT.}
    \label{Fig:cmp_abinit_mlff}
\end{figure}

Finally, we verify the accuracy of CNT-MLFF for the prediction of phonons. In particular, we choose the $(16,0)$ carbon nanotube for this study, which is not part of the training set, and consider a $2$-atom unit cell with $16 \, \nu_{\bq}$-points and $24 \, \eta_{\bq}$-points. We present the  the phonon frequencies and phonon density of states (DOS) so obtained in Fig.~\ref{Fig:cmp_abinit_mlff}.  The phonon DOS is calculated using the relation: 
\begin{align}
	g(\omega) = \frac{1}{ N_{\eta_\bq} \mathfrak{N}}  \sum_{\eta_\bq}^{}\sum_{\nu_\bq = 0}^{ \mathfrak{N}-1} \sum_{n=1}^{3\hN} \frac{\Delta}{\sqrt{2\pi}} \exp \left[-\left(\frac{\omega-\omega_n(\nu_\bq, \eta_\bq)}{\sqrt{2} \Delta}\right)^2 \right] \,,
\end{align} 
where $\hN$ is the number of atoms in the fundamental domain, $N_{\eta_\bq}$ is the number of wavevectors in the axial direction, $\mathfrak{N}$ is cyclic group order, and $\Delta$ is the Gaussian smearing width, which in the current case is chosen to be 0.25 cm$^{-1}$. To verify the accuracy of the CNT-MLFF model, we compare with results obtained by the established planewave DFT code ABINIT, while employing the density functional perturbation theory (DFPT) feature \citep{gonze2005brief}.  We choose a planewave cutoff of 80 Ha, which ensures that the reference phonon frequencies are converged to within $1$ cm$^{-1}$. Since ABINIT cannot exploit cyclic and helical symmetries, we consider an equivalent 64-atom periodic cell with $12 \, \eta_{\mathbf{q}}$-points. It is clear from the results that the phonon spectrum predicted by CNT-MLFF is in very good agreement with the first principles DFPT results, with the MAE, RMSE, and maximum error in the frequencies being $3.9$ cm$^{-1}$, $4.8$ cm$^{-1}$, and $12.9$ cm$^{-1}$, respectively, numbers that are commensurate with typically targeted numerical thresholds in DFPT based phonon calculations. 

%%%%%%%%%%%%%%%%%%%%%%%%%%%%%%%%%%%%%
\paragraph{\underline{Performance}} The CNT-MLFF model is highly efficient for the prediction of energies, atomic forces, and phonons. For instance, considering the relaxed (16,0) CNT, the total CPU time for prediction of the energy and forces is $\sim 0.1$ seconds, compared to $\sim 18$ hours for Cyclix-DFT and $\sim 240$ hours for ABINIT, which represents a speedup of  more than five orders-of-magnitude with respect to ab initio DFT calculations. Indeed, the computational time for Cyclix-DFT and therefore the speedup will reduce by a factor of $\sim 2.3$x when  the accuracy targeted in Cyclix-DFT is set to that achieved by CNT-MLFF. The total CPU time for prediction of the phonons using CNT-MLFF is $\sim 0.1$ hours, compared to $\sim 100,000$ hours in ABINIT, which represents a six orders-of-magnitude speedup relative to first principles DFPT calculations. Indeed, given the preliminary nature of the code for calculation of phonons in CNT-MLFF, the speedup is expected to remain at multiple orders-of-magnitude even when DFT based phonon calculations are implemented within the symmetry-adapted Cyclic-DFT formalism.  It is worth noting that due to symmetry-adaption, CNT-MLFF is expected to be generally more efficient relative to other MLFFs that have been developed for nanotubes \citep{vivanco2022machine, hedman2024dynamics, xiang2020machine, aghajamali2022superior}. 

In performing the above timing comparisons, we have neglected the training cost associated with the CNT-MLFF model. This is generally the case in the literature since training needs to be done only once during the lifetime of the model, i.e., the model can be repeatedly used for future simulations without incurring the cost associated with the training. In the current context, the CPU time associated with the training of the CNT-MLFF model is $\sim 30,000$ hours, which is even smaller than the time taken by ABINIT for the phonon calculation of the smallest nanotube.  Therefore, the cost associated with the training strategy adopted in this work can be considered to be negligible, demonstrating the efficacy of the training procedure adopted in this work. 

%%%%%%%%%%%%%%%%%%%%%%%%%%%%%%%%%%%%%%%%%%%%%%%%%%%%%%%%%%%%%%%%%%%%%%%%%

\subsection{Application: phonons in carbon nanotubes} \label{Sec:phononsinCNT}
\begin{figure}[htbp!]
    \centering
    \includegraphics[keepaspectratio=true,width=0.95\textwidth]{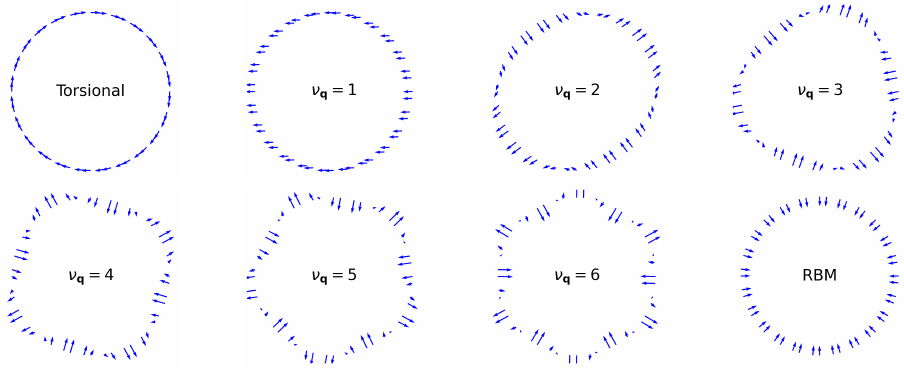}
    \caption{Representative phonon modes for CNTs.}
    \label{Fig:modes}
\end{figure}

We now use the trained CNT-MLFF model to study phonons in  CNTs of varying diameters and chiralities. As a representative set, we choose all $(n,0)$, $(n,n)$, $(2n,n)$, and $(3n,n)$ CNTs with  diameters between 1 and 2 nm. For obtaining the relaxed atomic configuratios, we perform  symmetry-adapted geometry relaxation with CNT-MLFF.  In the phonon calculations, we consider  a 2-atom unit cell with $5000$ $\eta_{\bq}$-points and the number of $\nu_{\bq}$-points equaling the cyclic group order.  In Fig.~\ref{Fig:modes}, we present the representative phonon modes that have been so predicted, namely rigid body modes, ring modes, and radial breathing modes (RBMs), which we now discuss.

 We observe that there are four zero-frequency rigid body phonon modes. These include three translational modes, which correspond to translations in Euclidean space, and one torsional mode, which corresponds to rotations about the nanotube's axis, arising due to the circular nature of the CNT cross-section. For all the CNTs, we obtain the longitudinal translational and torsional modes at the wavevector $(0,0)$, and the two transverse translational (flexure) modes at the phonon wavevectors $(1,\frac{\phi}{\tau})$ and $(-1,-\frac{\phi}{\tau})$, which is in agreement with the literature \citep{gunlycke2008lattice}. 
%%%%%%%%%%%%%%%%%%%%%%%%%%%%%%%%%%%%%%%%%%%%%%%%%%%%%%%%
%%%%%%%%%%%%%%%%%%%%%%%%%%%%%%%%%%%%%

\begin{figure}[htbp!]
    \centering
    \begin{subfigure}[b]{0.485\textwidth}
        \centering
        \includegraphics[keepaspectratio=true,width=0.97\textwidth]{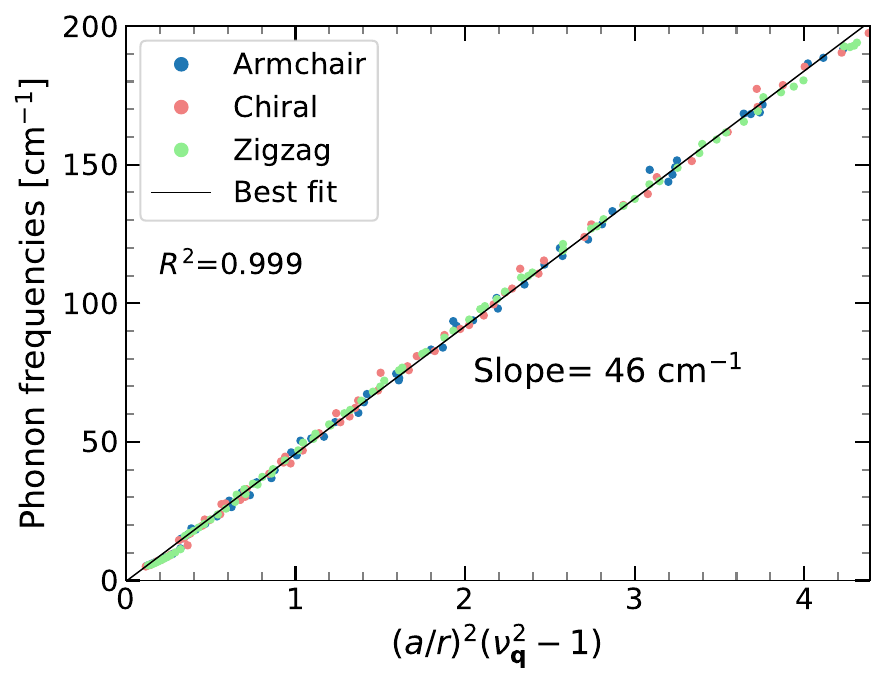}
        \caption{Ring modes}
        \label{fig:ring_fit}
    \end{subfigure}\hfill
    \begin{subfigure}[b]{0.49\textwidth}
        \centering
        \includegraphics[keepaspectratio=true,width=\textwidth]{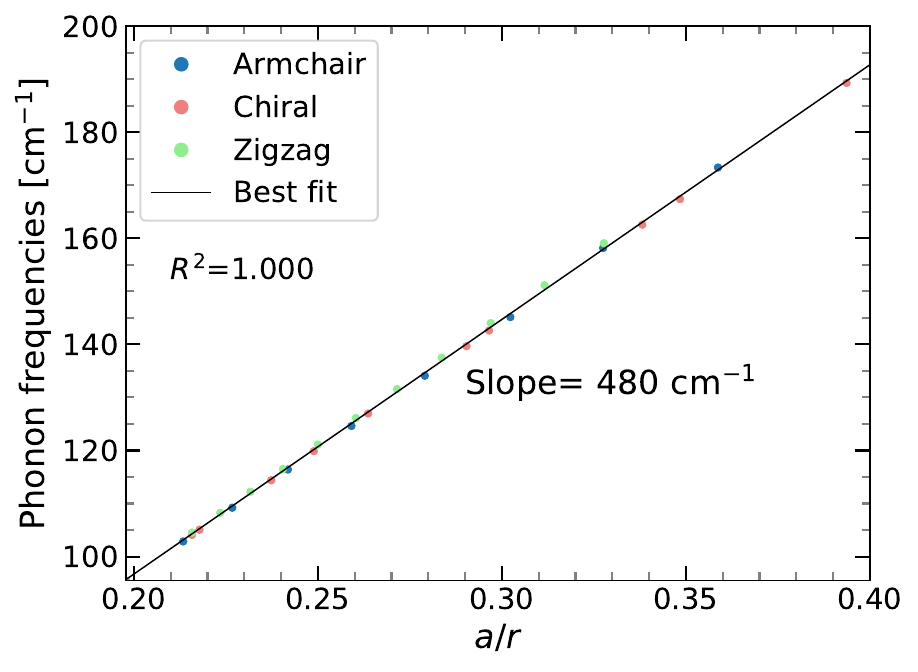}
        \caption{Radial breathing modes}
        \label{fig:RBM_fit}
    \end{subfigure}
    \caption{Phonon frequencies for the (a) ring modes and (b) radial breathing modes, when considering all zigzag, armchair, $(2n,n)$ chiral, and $(3n,n)$ chiral CNTs with diameters between 1 and 2 nm.}
    \label{Fig:modes_fit}
\end{figure}

We also observe the presence of low-frequency ring modes, wherein the atom vibrations form a ring-like pattern, as shown in Fig.~\ref{Fig:modes} for $\nu_{\bq}=1, 2, 3, 4, 5, 6$. These modes are represented by the symmetry present in them, e.g., $\nu_{\bq} = 3$ ring mode has 3-fold symmetry. For all the CNTs, we obtain the $\nu_{\bq}^\text{th}$ ring mode at the wavevector $(\nu_{\bq},\frac{\nu_{\bq} \phi}{\tau})$ of the lowest frequency branch. Furthermore, upon plotting the phonon frequencies of these modes for all diameters and chiralities, we observe that they only depend on the radius of the nanotube and the symmetry order of the mode (Fig.~\ref{fig:ring_fit}). This dependence can be expressed in terms of the following law:
\begin{align}
\omega_{\text{RM}} (r,\nu_\bq) \approx 46 \left(\frac{a}{r}\right)^2 (\nu_{\bq}^2-1) \, \text{cm}^{-1}\,,
\end{align}
where $a = \sqrt{3} a_0$ is the length of the graphene's lattice vector and $r$ is the radius of the nanotube.
%%%%%%%%%%%%%%%%%%%%%%%%%%%%%%%%%%%%%%%%%%%%%%%

We also observe the presence of radial breathing modes, where the atoms vibrate in the radial direction, causing an expansion and contraction of the nanotube. For all the CNTs, we obtain these modes at the wavevector $(0,0)$ corresponding to the third lowest phonon branch. Similar to the ring modes, the frequencies of these modes also depends only on the radius of the nanotube (Fig.~\ref{fig:RBM_fit}),  with the law given by:
\begin{align}
\omega_{\text{RBM}} (r) \approx 480 \frac{a}{r} \, \text{cm}^{-1}\,.
\end{align}

\begin{figure}[htbp!]
    \centering
    \begin{subfigure}[b]{0.49\textwidth}
        \centering
        \includegraphics[keepaspectratio=true,width=0.975\textwidth]{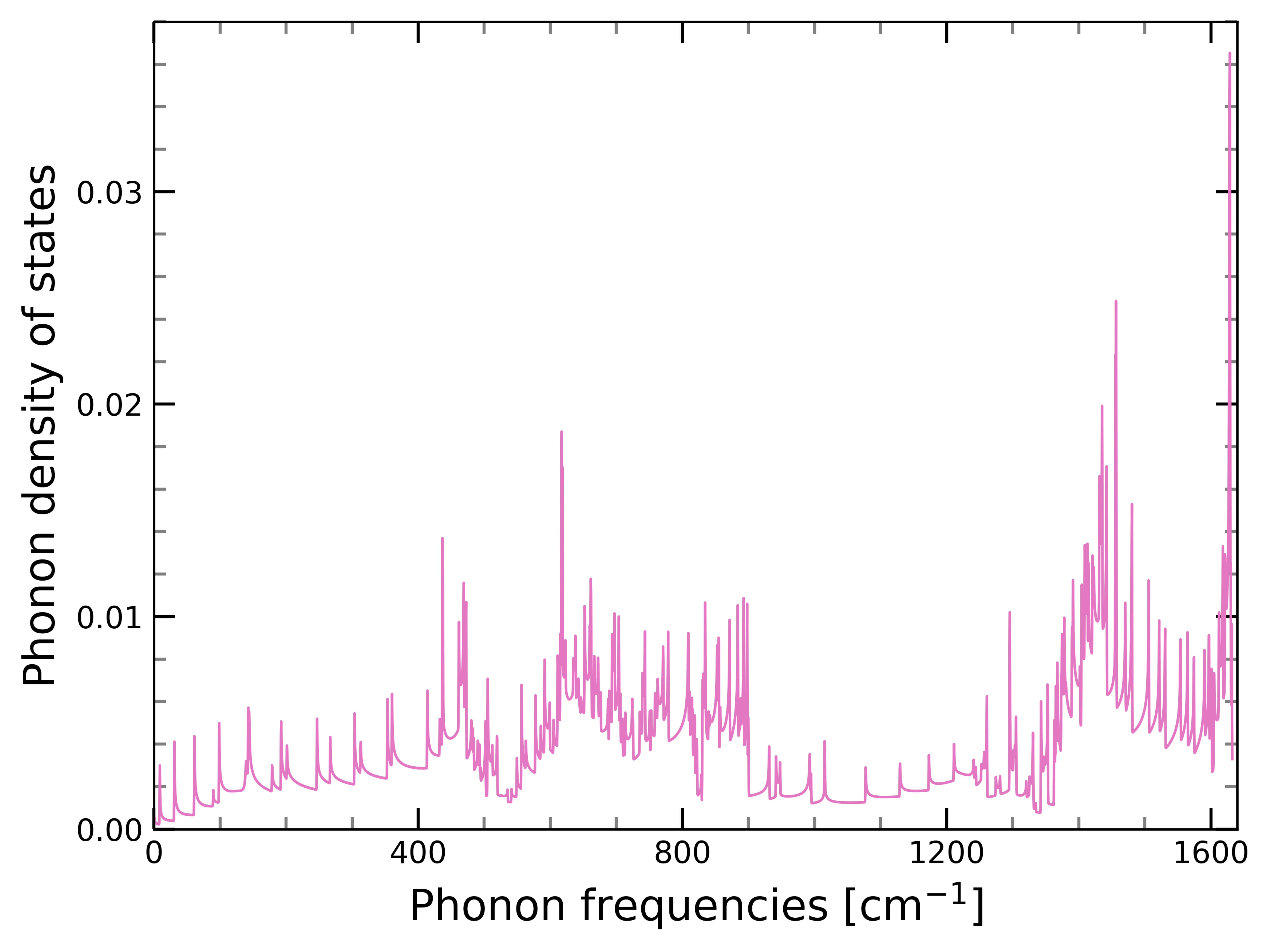}
        \caption{Phonon DOS}
        \label{fig:DOS_21_0_full}
    \end{subfigure}\hfill
    \begin{subfigure}[b]{0.49\textwidth}
        \centering
        \includegraphics[keepaspectratio=true,width=\textwidth]{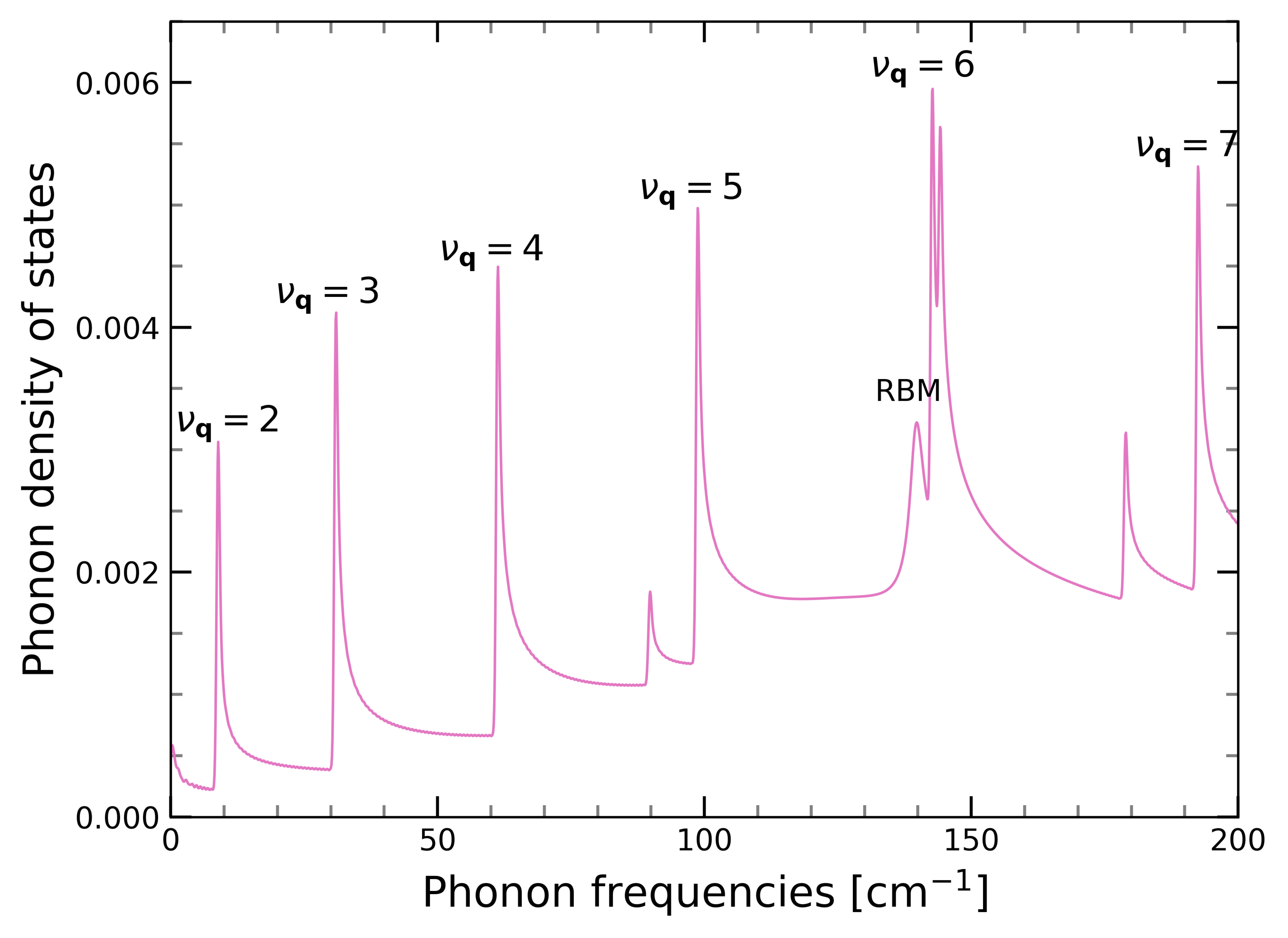}
        \caption{Phonon DOS at low frequencies}
        \label{fig:DOS_21_0_small}
    \end{subfigure}
    \caption{Phonon DOS for the zigzag $(21,0)$  CNT. (a) shows the DOS on the full spectrum, while (b) shows its low-frequency magnification. The ring modes and radial breathing mode are marked in (b) at the van Hove peaks.}
    \label{Fig:DOS_21_0}
\end{figure}

It is worth noting that the above phonon frequency laws for the ring and radial breathing modes  in CNTs are in very good agreement with those in literature \citep{kurti1998first, sanchez1999ab, mahan2004flexure, gunlycke2008lattice}. Indeed, the constants in these laws are different from those in the literature. The constants obtained in the current work are expected to be more precise, given the ab initio accuracy of the MLFF model. Furthermore, we obtain these important modes as van Hove singularities in the phonon DOS, as shown in Fig.~\ref{Fig:DOS_21_0}, which is also in agreement with the literature. This is a direct consequence of the quantum confinement effect, whereby the atoms are allowed to vibrate in discrete energy levels rather than continuous band, resulting in pronounced peaks in density of states.

%%%%%%%%%%%%%%%%%%%%%%%%%%%%%%%%%%%%%%%%%%%%
\section{Concluding remarks} \label{Sec:ConcludingRemarks}
In this work, we have presented a formalism for constructing cyclic and helical symmetry-informed MLFFs. In particular, using the SOAP descriptors in conjunction with the polynomial kernel method, we have derived cyclic and helical symmetry-adapted expressions for the energy, atomic forces, and phonons. We have utilized this formulation to develop a symmetry-informed MLFF for  single-walled CNTs, where the model was trained using Bayesian linear regression, with the data obtained from the first principles DFT calculations performed during on-the-fly symmetry-informed  MLFF MD simulations of representative CNTs. We have verified the accuracy of the symmetry-informed MLFF model by comparing its predictions of energies and forces with DFT calculations and its predictions of phonons with DFPT calculations for CNTs that were not included in the training. In particular, we have obtained a root mean square error of $1.4 \times 10^{-4}$ Ha/atom, $4.7 \times 10^{-4}$ Ha/Bohr, and 4.8 cm$^{-1}$ in the energy, forces, and phonon frequencies, respectively, which are well within the accuracy targeted in ab initio calculations. We have used this framework to study phonons in CNTs with different diameters and chiralities, through which we have identified a torsional rigid body mode specific to cylindrical structures, and established relationships for how the phonon frequencies of ring modes and radial breathing modes vary with the radius and chirality.

The proposed MLFF formalism provides an avenue for studying nanostructures with cyclic and helical symmetry at ab initio accuracy, while providing orders-of-magnitude speedup relative to such methods.  It also provides an avenue for investigating the effect of bending and torsional deformations in such systems, making it a worthy subject for future research. Extending the Cyclix-DFT framework to more advanced exchange-correlation functionals such as hybrid will further increase the range of systems that can be studied with quantum mechanical accuracy using MLFFs, making it another worthy subject for future research. 

%%%%%%%%%%%%%%%%%%%%%%%%%%%%%%%%%%%%%%%%%%%%%%%%%%%%%%%%%%%%%%
%%%%%%%%%%%%%%%%%%%%%%%%%%%%%%%%%%%%%%%%%%%%%%%%%%%%%%%%%%%%%%
	\section*{Acknowledgments}
	S.K and P.S. gratefully acknowledge the support of the U.S. Department of Energy, Office of Science under grant DE-SC0023445 for the development of the underlying on-the-fly MLFF MD framework. This research was also supported by the supercomputing infrastructure provided by Partnership for an Advanced Computing Environment (PACE) through its Phoenix cluster at Georgia Institute of Technology, Atlanta, Georgia.

%%%%%%%%%%%%%%%%%%%%%%%%%%%%%%%%%%%%%%%%%%%%%%%%%%%%%%%%%%%%%%
%%%%%%%%%%%%%%%%%%%%%%%%%%%%%%%%%%%%%%%%%%%%%%%%%%%%%%%%%%%%%%

\appendix

\section{Additional mathematical details}
\subsection{$\cnlm$ and its derivatives} \label{appendix:cnlm}
The function $\cnlm$ is defined as:
\begin{align} \label{eqn:cnlm}
\cnlm(\bRJI) = \hnl(|\bRJI|) \Ylm^*(\bRJI) \,,
\end{align}
where $\hnl$ is a radial function with compact support and $\Ylm$ is a spherical harmonic function. Its first order derivative can be calculated as:
\begin{align}\label{eqn:dcnlm}
\cnlm^{(1)}(\bRJI) = \frac{\bRJI}{|\bRJI|} \hnl^{(1)}(|\bRJI|) \Ylm^*(\bRJI) + \hnl(|\bRJI|) \Ylm^{*(1)}(\bRJI) \,,
\end{align}
where $f^{(1)}(\bx) = \frac{\partial f(\bx)}{\partial \bx}$. Similarly, its second order derivative can be calculated as:
\begin{align}\label{eqn:d2cnlm}
\cnlm^{(2)}(\bRJI) =& \left(\frac{\bI}{|\bRJI|}  - \frac{\bRJI}{|\bRJI|^3} (\bRJI)^{\rm T}\right) \hnl^{(1)}(|\bRJI|) \Ylm^*(\bRJI) \nonumber \\
&+ \frac{\bRJI}{|\bRJI|} \frac{(\bRJI)^{\rm T}}{|\bRJI|} \hnl^{(2)}(|\bRJI|) \Ylm^*(\bRJI) +  \hnl^{(1)}(|\bRJI|) \nonumber \\
&\times\Ylm^{*(1)}(\bRJI) \frac{(\bRJI)^{\rm T}}{|\bRJI|} + \frac{\bRJI}{|\bRJI|}  \hnl^{(1)}(|\bRJI|) \Ylm^{*(1)^{\rm T}}(\bRJI)\nonumber \\
&+ \hnl(|\bRJI|) \Ylm^{*(2)}(\bRJI) \,,
\end{align}
where $f^{(2)}(\bx) = \frac{\partial^2 f(\bx)}{\partial \bx \partial \bx^T}$, $\bI$ is the  identity matrix of size $3 \times 3$, and $(.)^{\rm T}$ denotes the transpose.

\subsection{Action of symmetry operator on spherical harmonics} \label{appendix:sphericalfunc_transform}
The spherical harmonic functions obey the following transformation under the action of the $\sfgm \in \sfG$ operator:
\begin{align}
\Ylm(\sfgm \circ (\bRJI)) &= \Ylm(\rotM (\bRJI)) \nonumber \\
&= \mathrm{e}^{i \m (\zeta \Theta +\mu \varphi)} \Ylm(\bRJI) \,,
\end{align}
where the phase factor in the last equality comes from the rotation of the spherical harmonic function in the azimuthal plane. The corresponding first and second order derivatives transform as follows:
\begin{align}
\Ylm^{(1)}(\sfgm \circ (\bRJI)) &= \mathrm{e}^{i \m (\zeta \Theta +\mu \varphi)} \rotM \Ylm^{(1)}(\bRJI) \,, \\
\Ylm^{(2)}(\sfgm \circ (\bRJI)) &= \mathrm{e}^{i \m (\zeta \Theta +\mu \varphi)} \rotM \Ylm^{(2)}(\bRJI) \rotMinv \,,
\end{align}
obtained by taking into account the action of the cyclic and helical symmetry group on the spherical harmonics and a vector in Euclidean space.

%%%%%%%%%%%%%%%
%\bibliography{Cyclix_MLFF}

\end{document}